\documentclass[]{interact}

\usepackage{epstopdf}
\usepackage[caption=false]{subfig}

\usepackage[numbers,sort&compress]{natbib}
\bibpunct[, ]{[}{]}{,}{n}{,}{,}
\renewcommand\bibfont{\fontsize{10}{12}\selectfont}
\makeatletter
\def\NAT@def@citea{\def\@citea{\NAT@separator}}
\makeatother

\theoremstyle{plain}

\theoremstyle{definition}

\theoremstyle{remark}

\begin{document}


\title{Quantum Wigner solid in two-dimensional electron systems in semiconductors}

\author{
\name{Alexander A. Shashkin\textsuperscript{a} and Sergey V. Kravchenko\textsuperscript{b}\thanks{CONTACT S.~V. Kravchenko. Email: s.kravchenko@northeastern.edu}}
\affil{\textsuperscript{a}Institute of Solid State Physics, Chernogolovka, Moscow District 142432, Russia\\
\textsuperscript{b}Physics Department, Northeastern University, Boston, MA 02115, USA}
}

\maketitle

\begin{abstract}
We review recent transport experiments that reveal two-threshold voltage-current characteristics, marked by a significant increase in noise between the two threshold voltages, at low electron densities in the insulating regime in two-dimensional (2D) electron systems, specifically in silicon metal-oxide-semiconductor field-effect transistors (MOSFETs) and SiGe/Si/SiGe heterostructures. The double-threshold voltage-current characteristics closely resemble those observed in the collective depinning of the vortex lattice in type-II superconductors. By adapting the model used for vortices to the case of an electron solid, good agreement with the experimental results is achieved, which supports a quantum electron solid forming in the low electron density state. When a perpendicular magnetic field is applied, the double-threshold behavior occurs at voltages an order of magnitude lower and at significantly higher electron densities than the zero-field case. This indicates the stabilization of the quantum electron solid, aligning with theoretical predictions. Interestingly, the double-threshold voltage-current curves, indicative of electron solid formation at low densities, are not observed in the quantum Hall regime. This lack of observation does not confirm the existence of a quasi-particle quantum Hall Wigner solid and indicates that quasi-particles near integer filling do not form an independent subsystem.
\end{abstract}


\newpage
\tableofcontents

\section{Introduction}

Experimental investigations into the transport and thermodynamic properties of two-dimensional (2D) electrons in semiconductors have indicated that these systems may approach a phase transition to an unknown state at low electron densities. This new state could be a quantum Wigner crystal or a precursor \cite{wigner1934on, chaplik1972possible, tanatar1989ground, shashkin2001indication, shashkin2002sharp,attaccalite2002correlation,spivak2004phases,shashkin2006pauli, mokashi2012critical,melnikov2017indication, kagalovsky2020hartree}. The term quantum refers to the fact that, in this context, the Fermi energy governs the kinetic energy of 2D electrons, in contrast to the classical Wigner crystal \cite{grimes1979evidence}, where the kinetic energy is determined by temperature.  The phase transition point in the least-disordered 2D electron systems in semiconductors is near the critical electron density for the metal-insulator transition, below which the 2D electrons become localized. Although the insulating regime of the metal-insulator transition has been extensively studied \cite{andrei1988observation,williams1991conduction,goldman1990evidence,jiang1990quantum,jiang1991magnetotransport,diorio1992reentrant,  pudalov1993zero,giamarchi2003electronic,qiu2012connecting,knighton2014reentrant,qiu2018new,knighton2018evidence, falson2022competing, hossain2022anisotropic,madathil2023moving}, there has been no definitive conclusion regarding the origin of the low-density state.  Single-threshold current-voltage ($I-V$) curves have been observed, which can be interpreted either as a sign of the depinning of an electron solid \cite{goldman1990evidence,williams1991conduction,diorio1992reentrant,pudalov1993zero} or as a breakdown of the insulating phase, consistent with traditional scenarios like strong electric field Efros-Shklovskii variable range hopping \cite{marianer1992effective} or percolation (see, \textit{e.g}., Refs.~\cite{jiang1991magnetotransport,dolgopolov1992metal,shashkin1994insulating,shashkin2005metal} and the references therein). Attempts to observe broadband voltage noise at the threshold of these $I-V$ curves and to probe the low-density state in perpendicular magnetic fields have not yielded sufficient information to clearly differentiate between the depinning of the electron solid and traditional mechanisms \cite{jiang1991magnetotransport,shashkin1994insulating,shashkin2005metal}. It is important to note that much confusion has arisen because many authors have chosen to interpret their data in the context of the Wigner crystal, overlooking conventional interpretations.

An important advance in uncovering the origin of the low-density state in strongly correlated 2D electron systems has been reported in Ref.~\cite{brussarski2018transport}. Two-threshold \( V-I \) characteristics have been observed, marked by a dramatic increase in noise between the two threshold voltages at the breakdown of the low-electron-density insulating phase. This noise manifests itself as strong current fluctuations; it rises significantly above the first threshold voltage \( V_{\text{th1}} \) and essentially disappears above the second threshold voltage \( V_{\text{th2}} \). The sharp noise peak in the \( V-I \) curves illustrates the double-threshold behavior. This phenomenon resembles the collective depinning of the vortex lattice in Type-II superconductors, with the voltage and current axes interchanged. The findings strongly favor the sliding 2D quantum electron solid, whereas the double-threshold behavior cannot be described within alternative scenarios such as percolation or overheating. It should be emphasized that, rather than being an ideal Wigner crystal, the 2D electron system studied is likely to be closer to an amorphous solid, similar to the vortex lattice in Type-II superconductors, where collective pinning has been well established. In previous studies, where the current was passed through the sample and the voltage between potential probes was measured, no distinct features were observed on almost flat curves in the breakdown regime. The main problem with those studies was that the noise in the voltage signal was small, which prevented the identification of a second threshold voltage.  In contrast, a DC voltage was applied between the source and the drain and the resulting induced current was measured. It was found that in this measurement configuration, the current noise is clear and dramatic. As a result, this experimental approach allowed the discovery of features that had not been previously observed.

Below, we describe transport experiments on Si MOSFETs and ultra-high-mobility SiGe/Si/SiGe heterostructures in both zero and perpendicular magnetic fields.

\section{Materials and Methods}\label{Methods}

Two sets of samples were utilized: silicon MOSFETs and SiGe/Si/SiGe heterostructures. Measurements were performed in an Oxford dilution refrigerator with a base temperature of approximately 30 mK. The direct current (DC) and noise were recorded using an ultra-low-noise current-voltage converter connected to a digital voltmeter or a lock-in amplifier.

The silicon (100) MOSFETs exhibited a peak electron mobility of approximately \(3 \, \text{m}^2 \, \text{V}^{-1} \, \text{s}^{-1}\) at temperatures below \(0.1 \, \text{K}\), which is similar to the values reported in Ref.~\cite{heemskerk1998nonlinear}. The electron density was controlled by applying a positive DC voltage to the gate in relation to the contacts, with an oxide thickness of \(150 \, \text{nm}\). The samples were designed with a Hall bar geometry, measuring \(50 \, \mu\text{m}\) in width.

Thin gaps were added to the gate metallization to tackle the primary experimental challenge of high contact resistance in the low-density, low-temperature regime. This design facilitated the maintenance of a high electron density near the contacts, independent of the density in the main section of the sample. As a result, contact resistances did not exceed approximately \(10 \, \text{k}\Omega\), allowing them to be disregarded in the insulating state. The voltage-current (\(V-I\)) characteristics and noise were measured in the central part of the sample, which had a length of \(180 \, \mu\text{m}\).

To create a SiGe/Si/SiGe sample, ultra-clean UHCVD-grown SiGe/Si/SiGe quantum wells (for details, see Refs.~\cite{lu2009observation,lu2010erratum,huang2012mobility}) were utilized. The electron mobility in these samples at low temperatures reached approximately 200 m$^2$/Vs, which is higher than that observed in previously used SiGe-based structures \cite{lu2009observation,lu2010erratum,huang2012mobility,schaffler1997high}. A silicon (001) quantum well, approximately 15 nm wide, was sandwiched between Si$_{0.8}$Ge$_{0.2}$ potential barriers \cite{melnikov2024triple}.  Contacts to the 2D layer were established using an approximately 300 nm thick Au$_{0.99}$Sb$_{0.01}$ alloy, which was deposited using a thermal evaporator and then annealed for 3-5 minutes at 440$^o$~C in an N$_2$ atmosphere. Through photolithography, the samples were patterned into Hall-bar shapes with a 150 $\mu$m distance between the potential probes and a width of 50 $\mu$m. An approximately 80 nm thick SiO layer was deposited on the wafer surface using a thermal evaporator, followed by depositing an approximately 40 nm thick aluminum gate on top of the SiO layer. Additionally, a roughly 20 nm thick layer of NiCr was applied on top of the aluminum to improve the adhesion of subsequent layers.  The contact gate was then fabricated, for which the structure was covered with an approximately 200 nm thick SiO layer, followed by an approximately 60 nm thick aluminum gate deposited on top of the SiO. The contact gate was designed to maintain a high electron density of about $2\times10^{11}$ cm$^{-2}$ near the contact areas, independent of the electron density in the main part of the sample. No intentional doping was applied; the electron density was controlled by applying a positive DC voltage to the gate in relation to the contacts. Saturating infrared illumination was applied to the samples for several minutes. This improved the quality of contacts and resulted in increased electron mobility \cite{melnikov2015ultra,melnikov2017unusual}. The resistance of the contacts remained below 10 k$\Omega$.

In addition to the double-gate samples, triple-gate samples were also used. The third gate was added to deplete the shunting channel between the contacts outside the Hall bar. This shunting channel can become prominent at the lowest electron densities in the insulating regime.  The triple gate samples had Hall bar shapes with a width of 50 $\mu$m and a distance between the potential probes of 100 $\mu$m. The main NiCr/Al Hall-bar gate was fabricated similarly to that in the double-gate samples, with the difference that an approximately 200~nm thick SiO layer was evaporated. The contact gate was made by covering the structure with a $\approx$150~nm thick SiO layer and depositing a $\approx$40~nm thick aluminum gate on top of the SiO.  An extra aluminum gate to deplete the shunting channel was fabricated simultaneously with the contact gate.  For more details, see Ref.~\cite{melnikov2024triple}.

\section{Results and Discussions}

\subsection{Double-threshold voltage-current curves as a signature of quantum electron solid; results obtained in Si MOSFETs}

\begin{figure}
\centering
\includegraphics[width=12 cm]{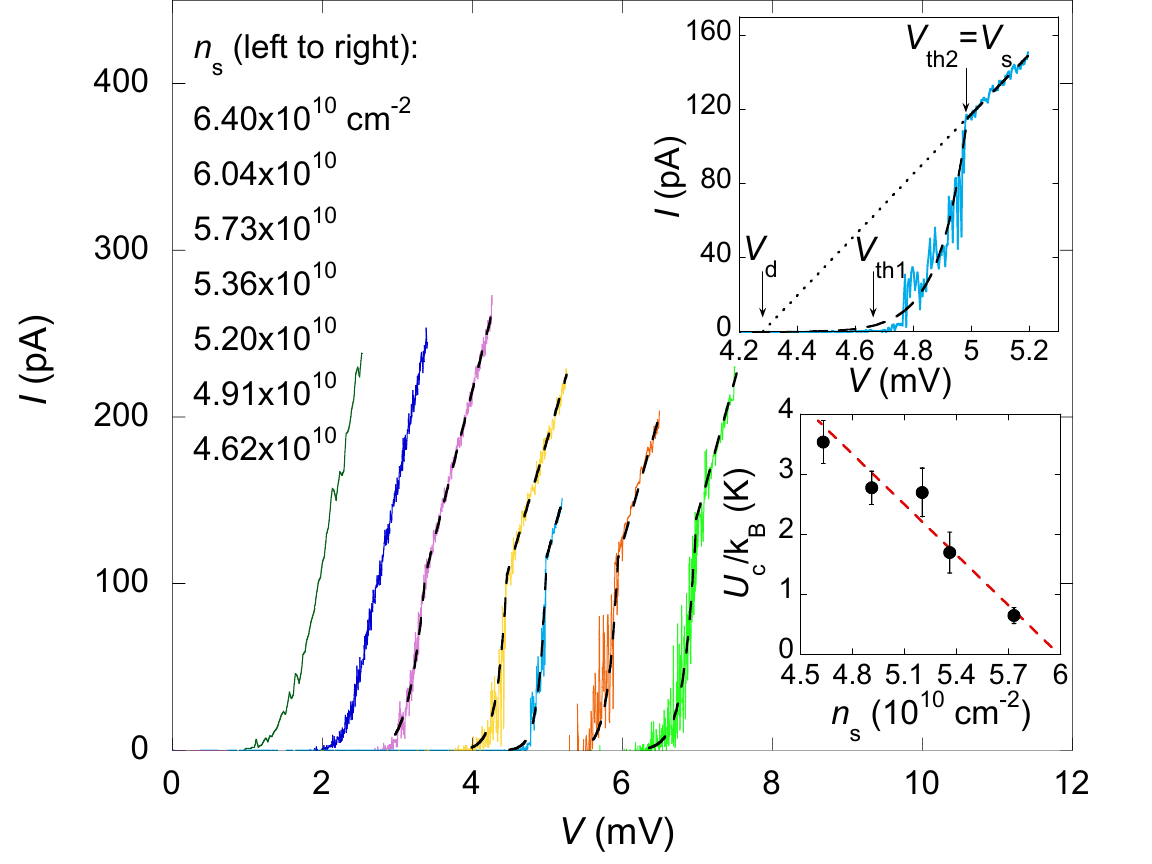}
\caption{$V-I$ characteristics measured at different electron densities in the insulating state at a temperature of 60~mK. The dashed lines are fits to the data using Eq.~(\ref{I}). In the top inset, the $V-I$ curve for $n_{\text{s}}=5.20\times 10^{10}$~cm$^{-2}$ is shown on an expanded scale; the threshold voltages $V_{\text{th1}}$ and $V_{\text{th2}}$, the static threshold $V_{\text{s}}=V_{\text{th2}}$, and the dynamic threshold $V_{\text{d}}$ (which is obtained by extrapolating the linear portion of the curves to $I=0$) are indicated. In the bottom inset, the activation energy $U_{\text{c}}$ \textsl{vs}.\ electron density is plotted. The dashed line is a linear fit.  Adapted from Ref.~\cite{brussarski2018transport}.}\label{fig1}
\end{figure}

Figure~\ref{fig1} illustrates a series of low-temperature voltage-current curves measured at different electron densities within the insulating regime, where \(n_{\text{s}} < n_{\text{c}}\). The critical density for the metal-insulator transition in this electron system is approximately \(n_{\text{c}} \approx 8 \times 10^{10} \, \text{cm}^{-2}\). The interaction parameter, which is defined as the ratio of the Coulomb energy to the Fermi energy, is given by \(r_{\text{s}} = g_{\text{v}} / (\pi n_{\text{s}})^{1/2} a_{\text{B}}\), where \(g_{\text{v}} = 2\) denotes the valley degeneracy, \(n_{\text{s}}\) is the areal density of electrons, and \(a_{\text{B}}\) is the effective Bohr radius in the semiconductor. At these densities, the value \(r_{\text{s}}\) exceeds \(r_{\text{s}} \sim 20\).  For electron densities below approximately \(6 \times 10^{10} \, \text{cm}^{-2}\), two threshold voltages are observed. As the applied voltage increases, the current remains near zero until the voltage reaches a first threshold, \(V_{\text{th1}}\). Once \(V_{\text{th1}}\) is surpassed, the current rises sharply until a second threshold voltage, \(V_{\text{th2}}\), is reached. Beyond \(V_{\text{th2}}\), the slope of the voltage-current curve decreases significantly, and the behavior becomes linear, although it is not ohmic (see also the top inset of Fig.~\ref{fig1}).  As the electron density is increased, the value of \(V_{\text{th1}}\) decreases, while the second threshold becomes less pronounced and eventually disappears. Notably, no hysteresis was observed within the range of electron densities studied.

\begin{figure}
\centering
\includegraphics[width=10 cm]{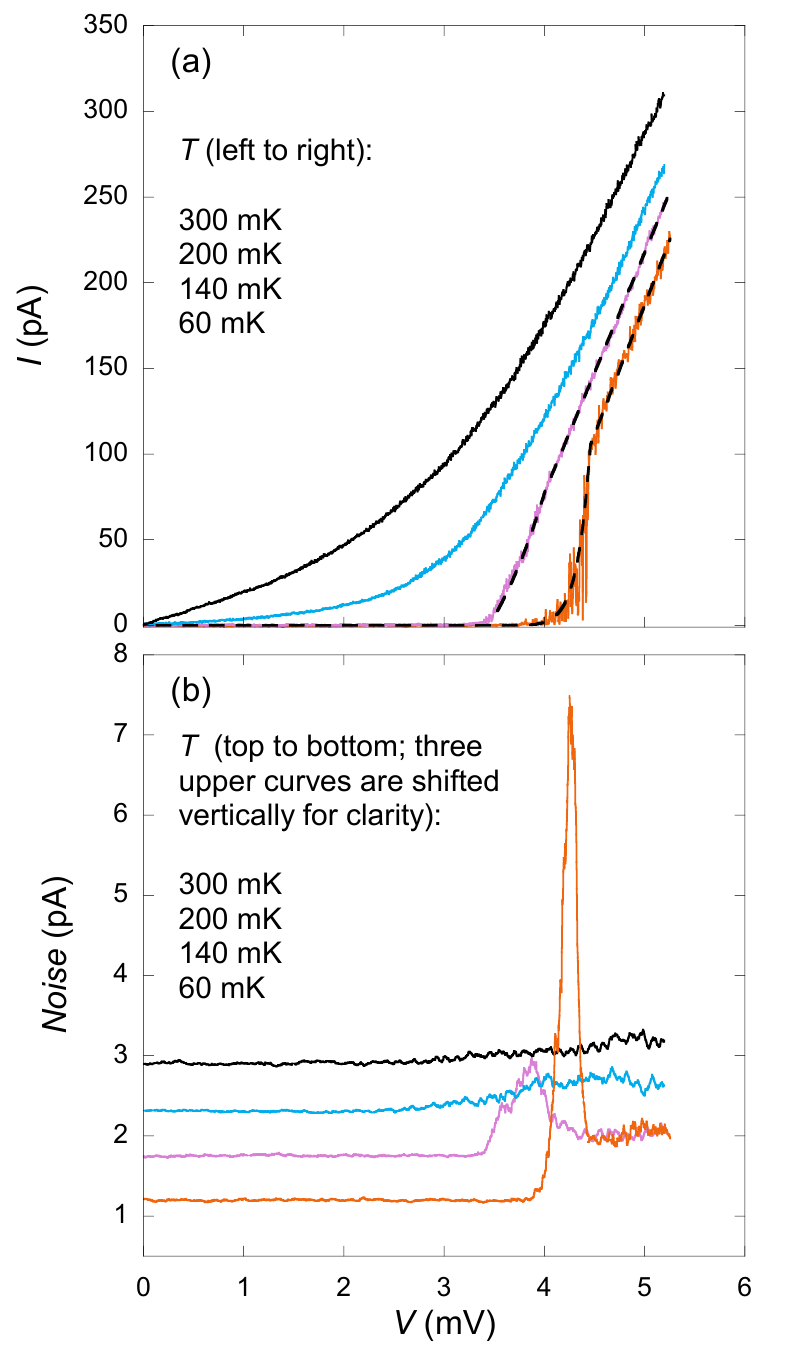}
\caption{(a) $V-I$ characteristics measured at $n_{\text{s}}=5.36\times 10^{10}$~cm$^{-2}$ and different temperatures. The dashed lines are fits to the data using Eq.~(\ref{I}). (b)~The broadband noise as a function of voltage for the same electron density and temperatures. Adapted from Ref.~\cite{brussarski2018transport}.}\label{fig2}
\end{figure}

Figure~\ref{fig2}(a) displays the $V-I$ characteristics for the electron density of $n_{\text{s}}=5.36\times 10^{10}$~cm$^{-2}$ at various temperatures. As the temperature $T$ increases, the second threshold voltage, $V_{\text{th2}}$, becomes less distinct, and the threshold behavior of the $V-I$ curves blurs.

The measured broadband noise is presented as a function of voltage in Fig. \ref{fig2}(b) for various temperatures, with an electron density of \( n_{\text{s}} = 5.36 \times 10^{10} \, \text{cm}^{-2} \). A significant increase in noise is observed between the thresholds \( V_{\text{th1}} \) and \( V_{\text{th2}} \) at the lowest temperature. This substantial noise decreases rapidly as the temperature increases, which is consistent with the two-threshold behavior seen in the \( V-I \) curves shown in Fig. \ref{fig2}(a).

\begin{figure}
\centering
\includegraphics[width=10 cm]{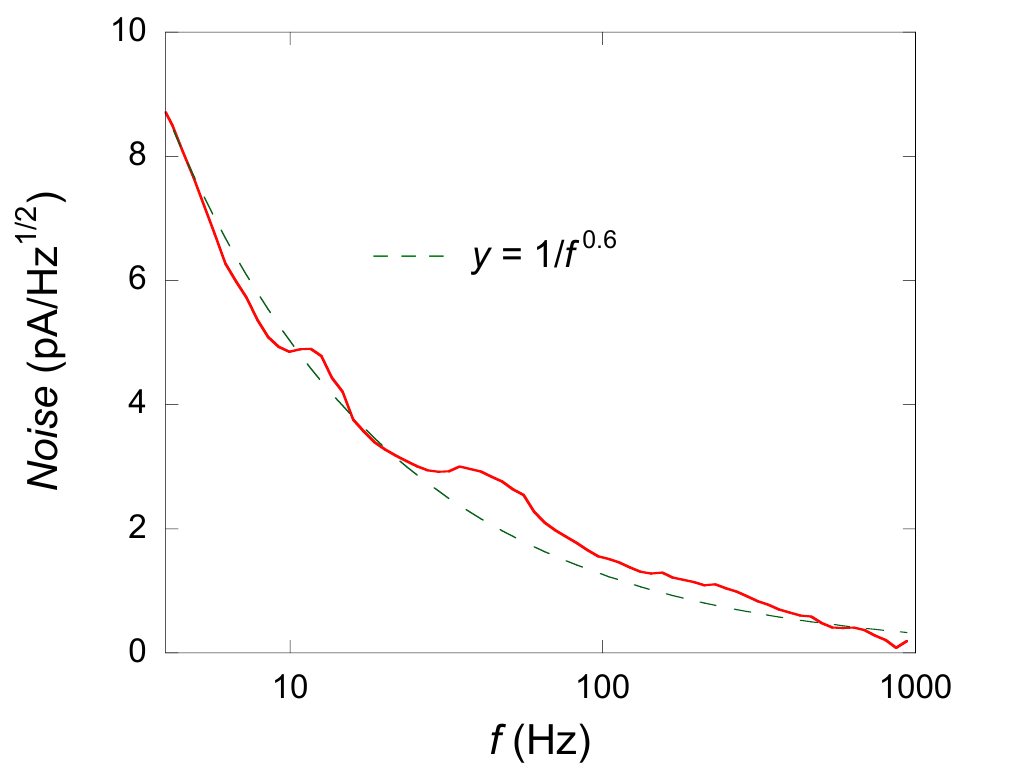}
\caption{Noise as a function of frequency at $n_{\text{s}}=5.36\times 10^{10}$~cm$^{-2}$, $T=60$~mK, and $V=4.26$~mV with resolution/bandwidth 2.5~Hz. The broad maxima at $f\sim10$ and 60~Hz on the order of 1 pA/$\sqrt{{\text{Hz}}}$ are within the experimental uncertainty. The dashed line shows the $1/f^{0.6}$ dependence.  Adapted from  Ref.~\cite{brussarski2018transport}.}\label{fig3}
\end{figure}

The spectrum of the generated noise, measured at its maximum value, is shown in Fig.~\ref{fig3}. The noise level increases as the frequency, \( f \), decreases, following the \( 1/f^\alpha \) law with \( \alpha \approx 0.6 \), which is close to one. This observation aligns with previous findings that indicated noise of the form \( 1/f^\alpha \) with \( \alpha \) near one in the linear response regime of similar samples close to the metal-insulator transition \cite{jaroszynski2002universal,jaroszynski2004magnetic}.

There is a striking similarity between the double-threshold $V-I$ dependences in the insulating state of Si MOSFETs and those (with the voltage and current axes interchanged) known for the depinning of the vortex lattice in type-II superconductors (see, for example, Refs.~\cite{yeh1991flux,blatter1994vortices,bullard2008vortex}). The physics of the vortex lattice in type-II superconductors, in which the existence of two thresholds is well known, can be adapted for the case of an electron solid. The transient region between the dynamic ($V_{\text{d}}$) and static ($V_{\text{s}}$) thresholds corresponds to the collective pinning of the solid. In this region, the pinning occurs at the centers with different energies, and the current is thermally activated:

\begin{equation}
I\propto\exp\left[-\frac{U(V)}{k_{\text{B}}T}\right],
\label{exp}
\end{equation}
where $U(V)$ is the activation energy. The static threshold $V_{\text{s}}$ signals the onset of the regime of
solid motion with friction. This corresponds to the condition

\begin{equation}
eEL=U_{\text{c}},
\label{Uc}
\end{equation}
where $E$ is the electric field and $L$ is the characteristic distance between the pinning centers with maximal activation energy $U_{\text{c}}$. From the balance of the electric, pinning, and friction forces in the regime of solid motion with friction, one expects a linear $V-I$ characteristic that is offset by the threshold $V_{\text{d}}$ corresponding to the pinning force

\begin{equation}
I=\sigma_0(V-V_{\text{d}}),
\label{linear}
\end{equation}
where $\sigma_0$ is a coefficient. Assuming that the activation energy for the Wigner solid is equal to

\begin{equation}
U(V)=U_{\text{c}}-eEL=U_{\text{c}}(1-V/V_{\text{s}}),
\label{U}
\end{equation}
the expression for the current is obtained:

\begin{equation}
I=\left\{\begin{array}{c}
\sigma_0(V-V_{\text{d}}) {\text{ if }} V>V_{\text{s}}\\
\sigma_0(V-V_{\text{d}})\exp\left[-\frac{U_{\text{c}}(1-V/V_{\text{s}})}{k_{\text{B}}T}\right] {\text{ if }} V\leq V_{\text{s}}.
\end{array}\right.\label{I}
\end{equation}
The fits to the data using Eq.~(\ref{I}) are shown by dashed lines in Figs.~\ref{fig1} and \ref{fig2}(a). As the figures show, the experimental two-threshold $V-I$ characteristics are described well by Eq.~(\ref{I}). The value of $U_{\text{c}}$ decreases approximately linearly with electron density and tends to zero at $n_{\text{s}}\approx 6\times 10^{10}$~cm$^{-2}$ (the bottom inset of Fig.~\ref{fig1}). This is in contrast to the vanishing activation energy of electron-hole pairs at $n_{\text c}$ obtained by measurements of the resistance in the limit of zero voltages/currents \cite{shashkin2001metal}. Presumably, the vanishing $U_{\text{c}}$ is related to the minimum number of the strong pinning centers for which the collective pinning is still possible. The fact that the coefficient $\sigma_0$ is approximately constant ($\sigma_0\approx 1.6\times 10^{-7}$~Ohm$^{-1}$) indicates that weak pinning centers control the solid motion with friction \cite{blatter1994vortices}. Thus, the physics of the vortex lattice, adapted for the case of an electron solid, is relevant for the insulating state in a 2D electron system in silicon.

\subsection{Generality of the double-threshold voltage-current curves for different classes of electron systems; results obtained in SiGe/Si/SiGe heterostructures}

As inferred from both the level and character of the disorder potential, Si MOSFETs and unprecedentedly high-mobility heterostructures, including SiGe/Si/SiGe heterostructures, belong to different classes of electron systems.

\begin{figure}
\centering
\includegraphics[width=8 cm]{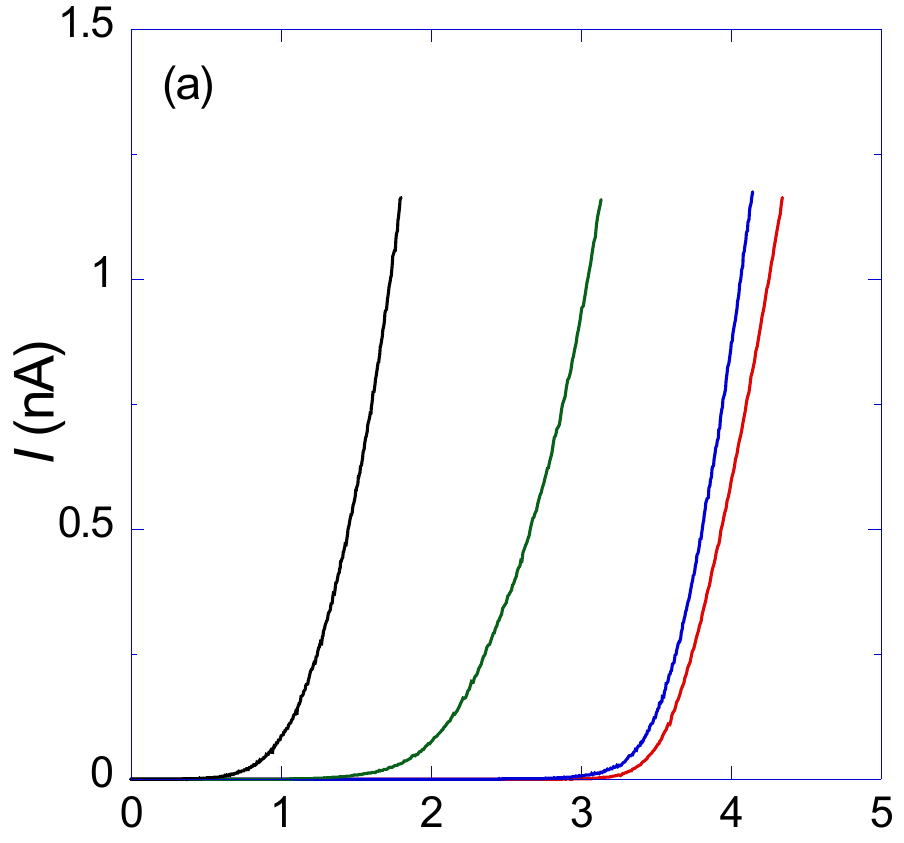}\\
\includegraphics[width=8 cm]{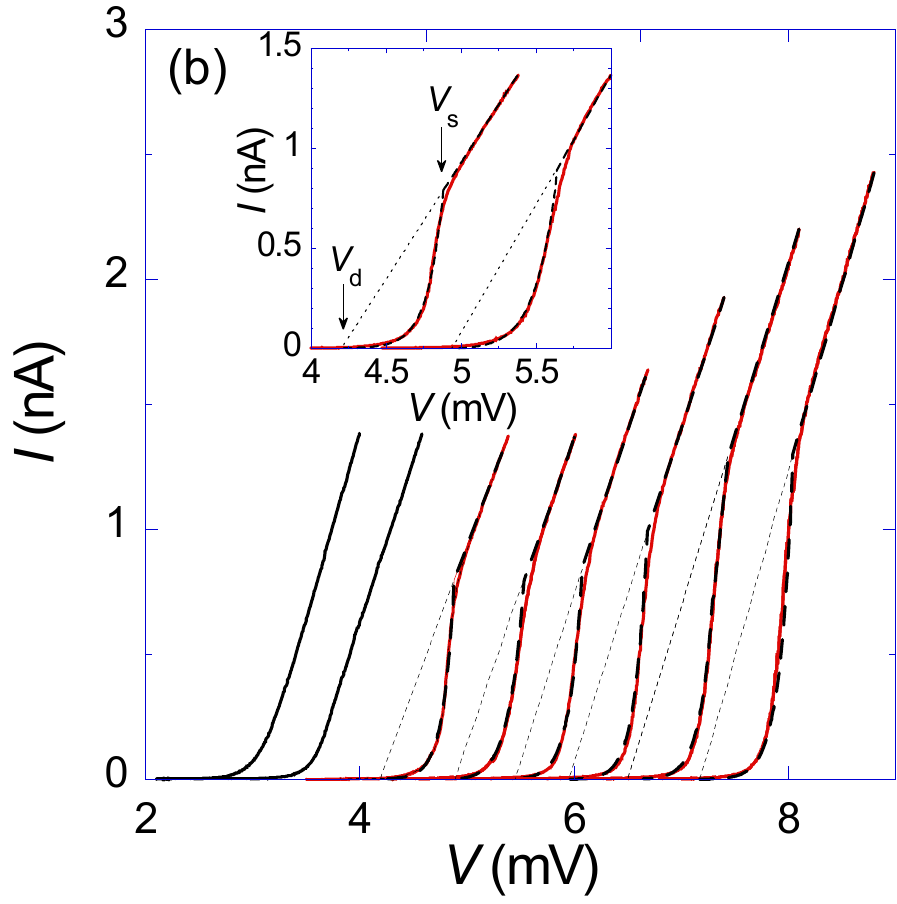}
\caption{(a) The $V$-$I$ characteristics measured in double-gate samples at various electron densities, displayed from left to right as follows: 7.95, 6.98, 6.21, and 5.82 $\times 10^9$ cm$^{-2}$. The temperature is $T = 30$ mK. In panel (b), the $V$-$I$ characteristics for triple-gate samples are shown at different electron densities, listed from left to right as: 6.37, 6.19, 6.01, 5.92, 5.83, 5.74, 5.65, and 5.56 $\times 10^9$ cm$^{-2}$, also at $T = 30$ mK. The dashed lines represent fits to the data using Equation ({\ref I}). The inset displays the $V$-$I$ characteristics for electron densities of $n_{\text s} = 6.01 \times 10^9$ cm$^{-2}$ and $5.92 \times 10^9$ cm$^{-2}$ on an expanded scale. Arrows indicate the dynamic threshold voltage $V_{\text d}$, which is obtained by extrapolating (as shown by the dotted line) the linear portion of the $V$-$I$ curves to zero current, along with the static threshold voltage $V_{\text s}$. From Ref.~\cite{melnikov2024triple}.}\label{fig4}
\end{figure}

In Fig.~\ref{fig4}(a), the $V$-$I$ characteristics measured in the insulating regime at low electron densities ($r_{\text{s}}>20$) in double-gate samples are presented.  With increasing applied voltage, the current remains near zero up to a certain threshold voltage; beyond this threshold, the current sharply increases, the threshold voltage increasing as the electron density decreases.  However, these single-threshold $V$-$I$ characteristics stop changing below $n_{\text s}\approx6\times10^9$~cm$^{-2}$. This indicates the presence of a shunting conduction channel outside the Hall bar that is obviously related to residual unintentional donor impurities in the SiGe/Si/SiGe heterostructures.

Applying a negative voltage to the additional gate in triple-gate samples depletes the shunting channel. However, the total depletion has not been reached, staying in the parallel-plate capacitor regime, in which the additional gate does not influence the transport properties of the 2D electron system in the main part of the sample. By suppressing the shunting channel, a significantly different behavior in the voltage-current characteristics is observed as the electron density decreases, as illustrated in Fig.~\ref{fig4}(b). The $V-I$ characteristics show variations across the entire range of electron densities studied when the electron density is decreased. At the lowest electron densities, a second threshold voltage appears in the $V-I$ curves. These characteristics resemble those previously observed in Si MOSFETs \cite{brussarski2018transport}. This shows that the strongly interacting limit at low electron densities can be realized in the 2D electron system in SiGe/Si/SiGe heterostructures.

\begin{figure*}
\centering
\includegraphics[width=7 cm]{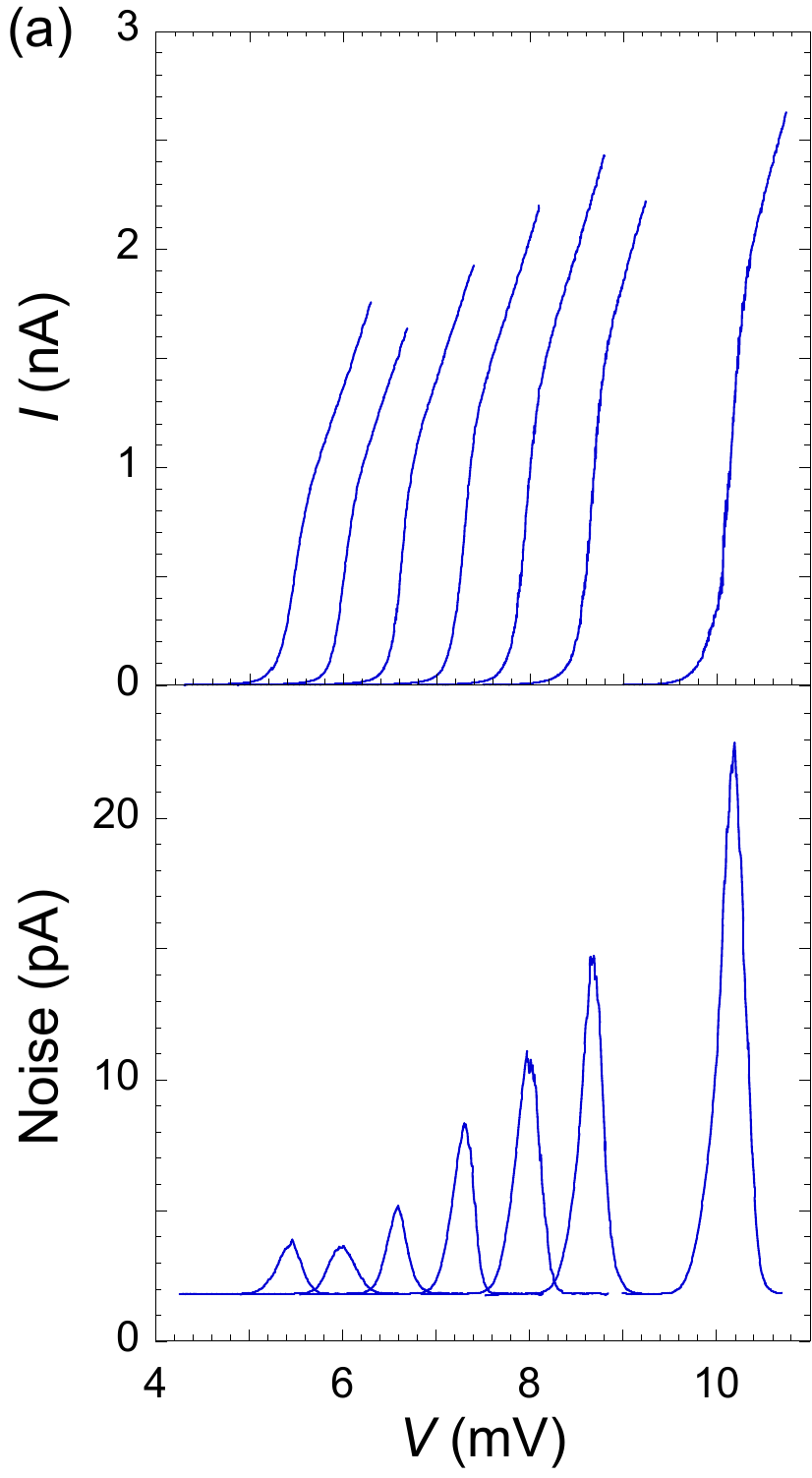}
\includegraphics[width=7 cm]{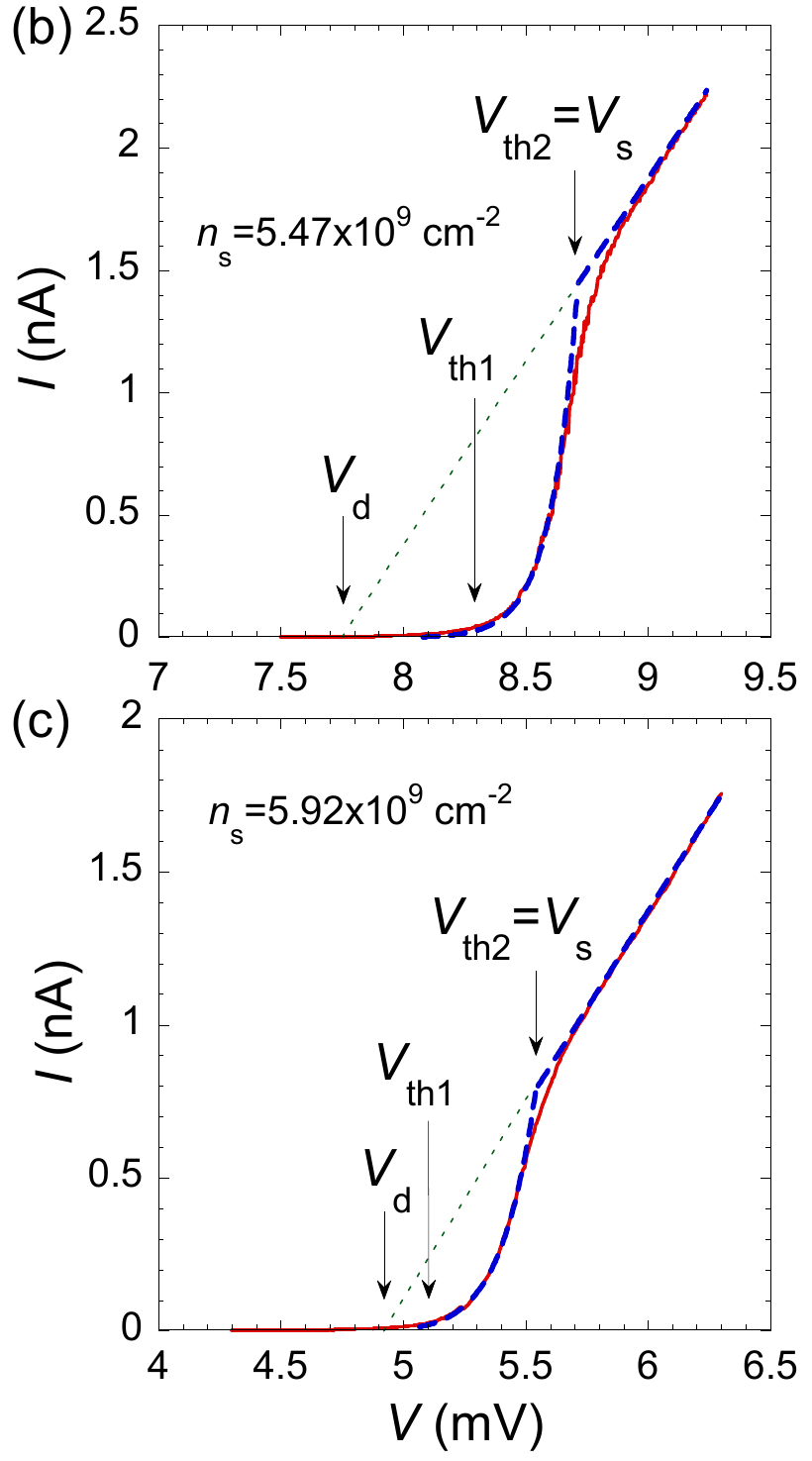}
\caption{(a) The $V$-$I$ characteristics are shown in the upper panel and the broadband noise is shown in the lower panel at a temperature of approximately 30~mK and various electron densities (from left to right): 5.92, 5.83, 5.74, 5.65, 5.56, 5.47, and $5.29\times10^9$~cm$^{-2}$.  Panels (b) and (c) display the $V$-$I$ characteristics for two electron densities on an expanded scale.  Also indicated are the threshold voltages, $V_{\text{th1}}$ and $V_{\text{th2}}$, the dynamic threshold $V_{\text d}$, which is obtained by extrapolating the linear portion of the $V$-$I$ curves to zero current, and the static threshold $V_{\text s}=V_{\text{th2}}$.  The dashed lines are fits to the data using Eq.~(\ref{I}). From Ref.~\cite{melnikov2024collective}.}\label{fig5}
\end{figure*}

In Fig.~\ref{fig5}(a), the typical voltage-current characteristics (upper panel) and the generated noise (lower panel) measured at a temperature of approximately 30 mK are presented. These measurements were conducted across different electron densities within the insulating regime (see also Fig.~\ref{fig5}(b, c) for a
more detailed view) \cite{melnikov2024collective}. These densities are below, but not too close to, the critical density \( n_{\text c} \approx 8.8 \times 10^9 \, \text{cm}^{-2} \), which signifies the transition from a metallic to an insulating state in the samples studied. At these specific values of \( n_{\rm s} \), the interaction parameter \( r_{\text s} \) exceeds 20. Notably, between the two threshold voltages, a peak in broadband current noise is observed, as shown in the lower panel of Fig.~\ref{fig5}(a).

\begin{figure}
\centering
\includegraphics[width=7 cm]{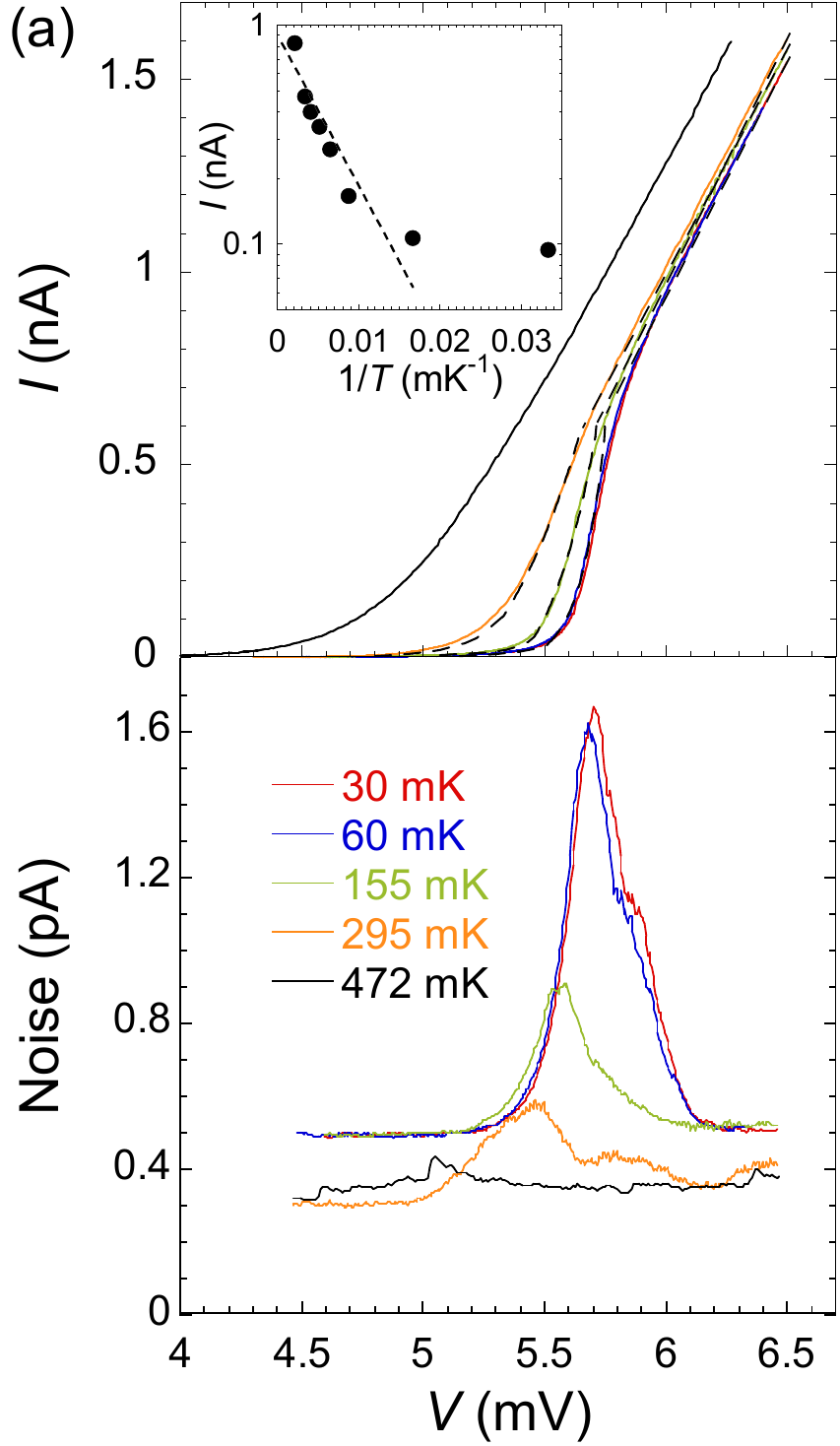}
\includegraphics[width=7 cm]{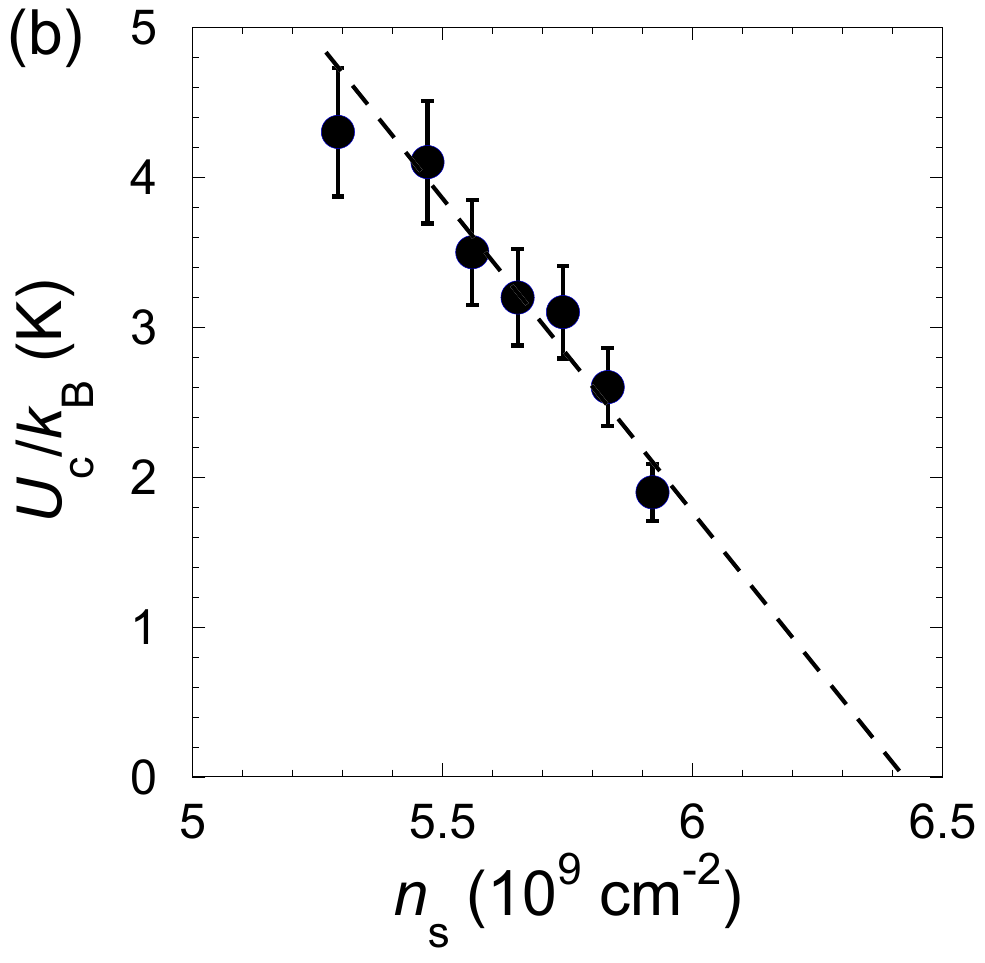}
\caption{(a) The current (upper panel) and broadband noise (lower panel) are shown as a function of voltage at an electron density of \(n_{\text{s}} = 5.83 \times 10^9 \, \text{cm}^{-2}\) for different temperatures in another sample. The curves in the upper panel are color-coded to correspond with the same temperatures presented in the lower panel. The overall noise is measured over the same frequency range as depicted in Fig.~\ref{fig7}. The dashed lines represent fits to the data using Eq.~(\ref{I}). The inset includes an Arrhenius plot of \(I(T)\) at \(n_{\text{s}} = 5.83 \times 10^9 \, \text{cm}^{-2}\) and \(V = 5.6 \, \text{mV}\). The dashed line is a linear fit that excludes the data point at 30 mK. (b) The activation energy \(U_{\text{c}}\) is plotted as a function of the electron density in the sample from Fig.~\ref{fig5}. The dashed line indicates a linear fit. From Ref.~\cite{melnikov2024collective}.}\label{fig6}
\end{figure}

Figure~\ref{fig6}(a) shows how the voltage-current characteristics and noise levels change with temperature. As the temperature rises, the voltage-current characteristics become less steep between the two threshold voltages, and the curves shift to lower voltages until the behavior associated with these two thresholds disappears. Similarly, the noise peak diminishes and eventually disappears as temperature increases. The inset of the upper panel in Fig.~\ref{fig6}(a) features an Arrhenius plot of \( I(T) \), which illustrates the activation temperature dependence down to approximately 60 mK. However, the data point collected at the lowest temperature deviates from this expected activation behavior, despite the electron temperature reached approximately 30 mK, as inferred from the analysis of Shubnikov-de~Haas oscillations in the metallic regime. This deviation can be attributed to residual sample inhomogeneities, although the possible overheating effects in this experiment cannot be completely discounted.

\begin{figure}
\centering
\includegraphics[width=9 cm]{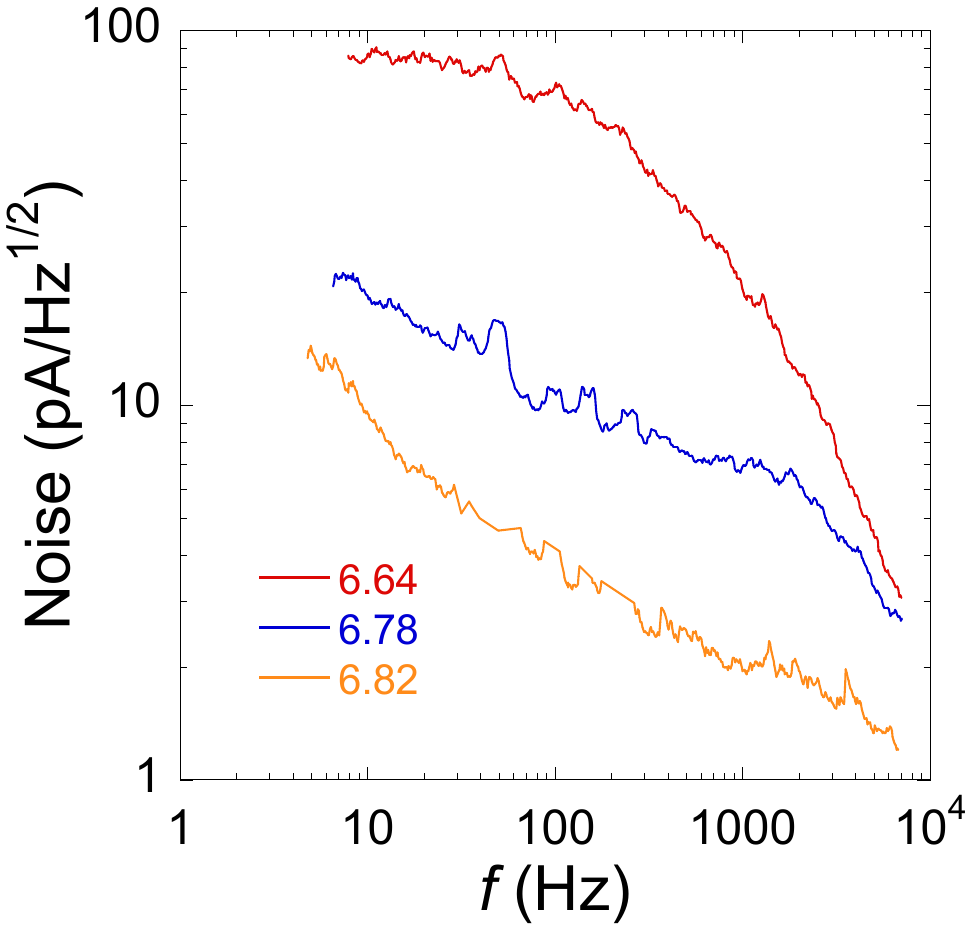}
\caption{The frequency dependence of noise at the maximum at three electron densities indicated in units of $10^9$~cm$^{-2}$ and at $T\approx30$~mK in another sample. From Ref.~\cite{melnikov2024collective}.}\label{fig7}
\end{figure}

In Fig.~\ref{fig7}, the noise spectra at the maximum of noise measured at three different electron densities and a temperature of approximately 30 mK are presented. At high frequencies, the noise shows a more pronounced dependence on frequency as the electron density decreases, approaching a \( 1/f^\alpha \) dependence with \( \alpha \approx 1 \) for the two lowest densities. As the frequency decreases, the behavior at these two lowest densities changes to \( 1/f^\alpha \) with \( \alpha \approx 0.2 \), which is similar to the dependence observed at the highest electron density. This change in \( \alpha \) occurs at a frequency that decreases with decreasing electron density.

Similarly to silicon MOSFETs, the observed results can be explained through a phenomenological theory of the collective depinning of elastic structures. This naturally produces a peak in broadband current noise between the dynamic and static thresholds, transitioning to the sliding of the solid over a pinning barrier once the static threshold is exceeded. The findings provide evidence for the formation of a quantum electron solid within this electron system and demonstrate the generality of this effect across different classes of electron systems.

\subsection{Stabilization of the quantum electron solid in perpendicular magnetic fields in SiGe/Si/SiGe heterostructures}

\begin{figure}
\centering
\includegraphics[width=7 cm]{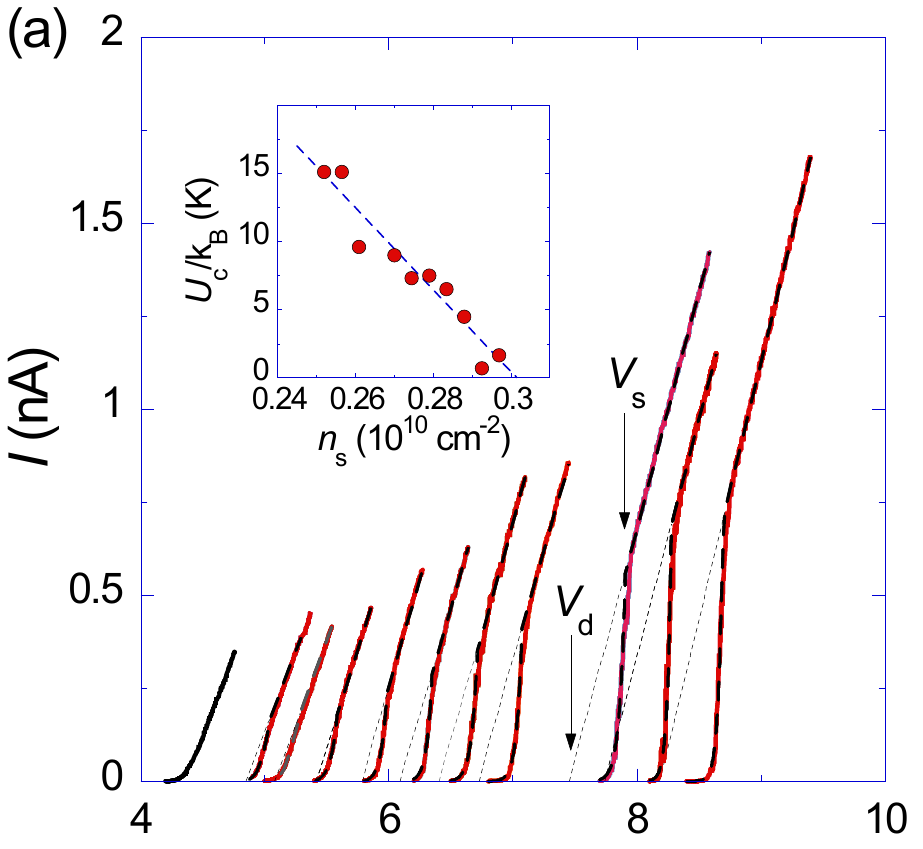}\\
\includegraphics[width=7 cm]{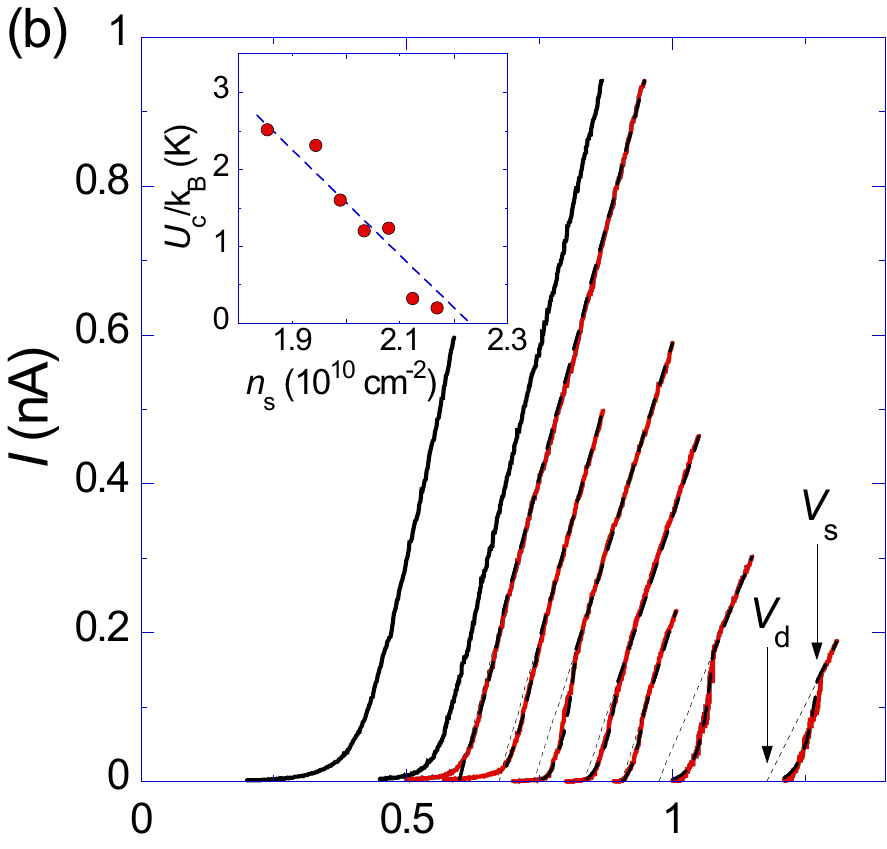}\\
\includegraphics[width=7 cm]{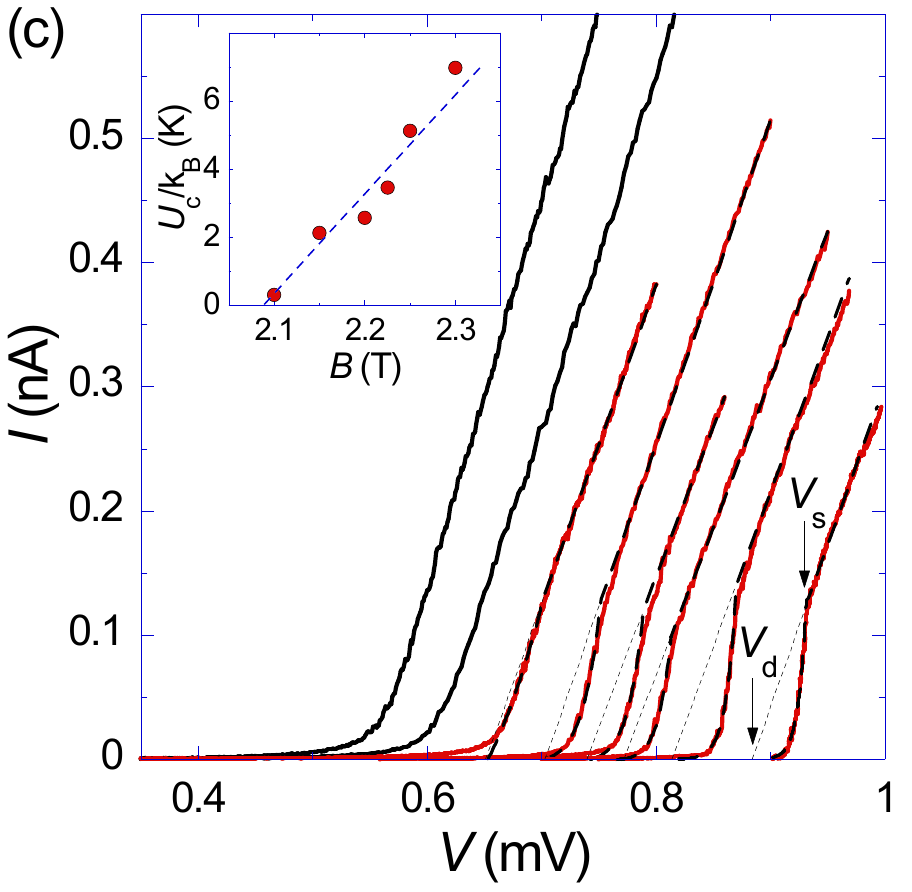}
\caption{(a) Voltage-current characteristics in zero magnetic field at $T=60$~mK at electron densities (in units of $10^{10}$~cm$^{-2}$, left to right): 0.306, 0.297, 0.293, 0.288, 0.284, 0.279, 0.275, 0.270, 0.261, 0.257, 0.252.  (b)~Voltage-current characteristics in $B=3$~T at $T=60$~mK at electron densities (in units of $10^{10}$~cm$^{-2}$, left to right): 2.30, 2.21, 2.17, 2.12, 2.08, 2.03, 1.99, 1.94, 1.85.  (c)~Voltage-current characteristics at $T=60$~mK for $n_{\text{s}}=1.67\times 10^{10}$~cm$^{-2}$ in different magnetic fields (left to right): 2, 2.05, 2.1, 2.15, 2.2, 2.225, 2.25, and 2.3~T.  Also shown are the dynamic threshold $V_{\text d}$ obtained by the extrapolation (dotted line) of the linear part of the $V$-$I$ curves to zero current and the static threshold $V_{\text s}$.  The dashed lines are fits to the data using Eq.~({\ref I}).  Insets: activation energy $U_{\text c}$ as a function of the electron density in (a, b) and the magnetic field in (c).  The dashed line is a linear fit. From Ref.~\cite{melnikov2025stabilization}.}\label{fig8}
\end{figure}

In Fig.~\ref{fig8}(a), a set of voltage-current characteristics is displayed.  The curves are measured at a temperature of 60~mK in zero magnetic field at different electron densities in the insulating regime $n_{\text{s}}<n_{\text{c}}$ (here $n_{\text{c}}\approx 0.7\times 10^{10}$~cm$^{-2}$ is the critical density for the metal-insulator transition in the samples used); the corresponding interaction parameter $r_{\text{s}}$ exceeds 20 at these values of $n_{\text{s}}$.  Two-threshold voltage-current curves are observed at electron densities below $n_{\text{s}}\approx 0.3\times 10^{10}$~cm$^{-2}$ \cite{melnikov2025stabilization}.

In Fig.~\ref{fig8}(b), $V-I$ curves at a temperature of \( T = 60 \) mK for \( B = 3 \) T are shown. The figure shows that the double-threshold behavior occurs at voltages that are an order of magnitude lower and at significantly higher electron densities compared to the case without a magnetic field. A peak in broadband current noise is observed between the two threshold voltages, whether or not a magnetic field is present (not shown here).

\begin{figure}
\centering
\includegraphics[width=8 cm]{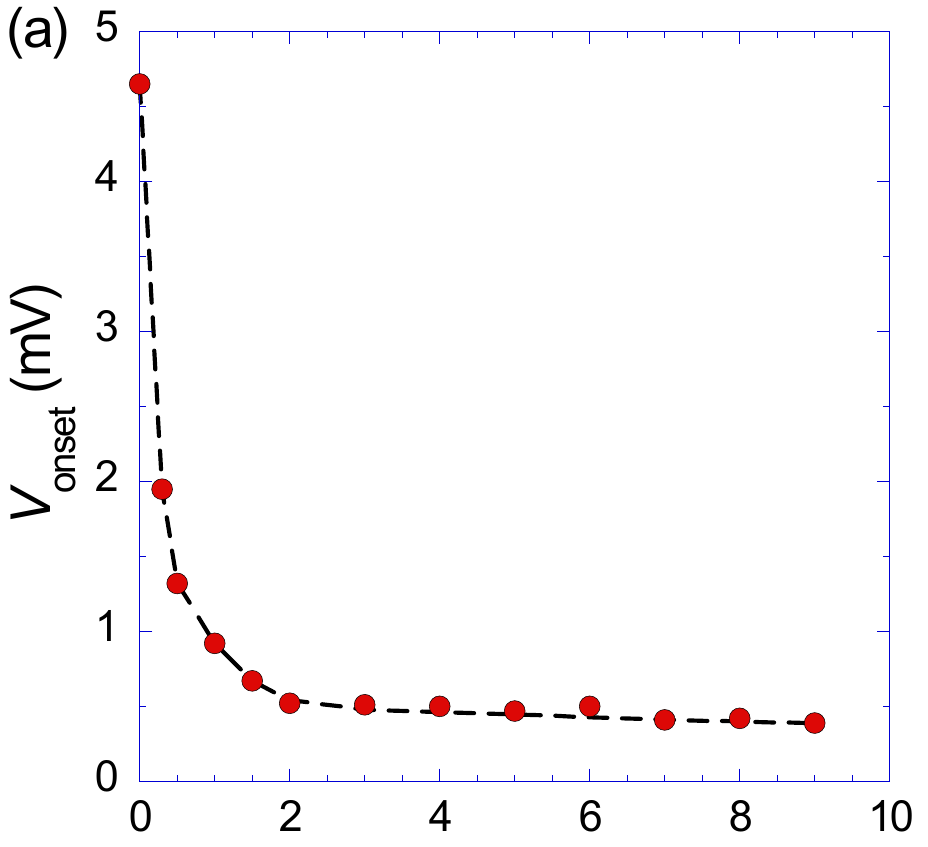}\\
\includegraphics[width=8 cm]{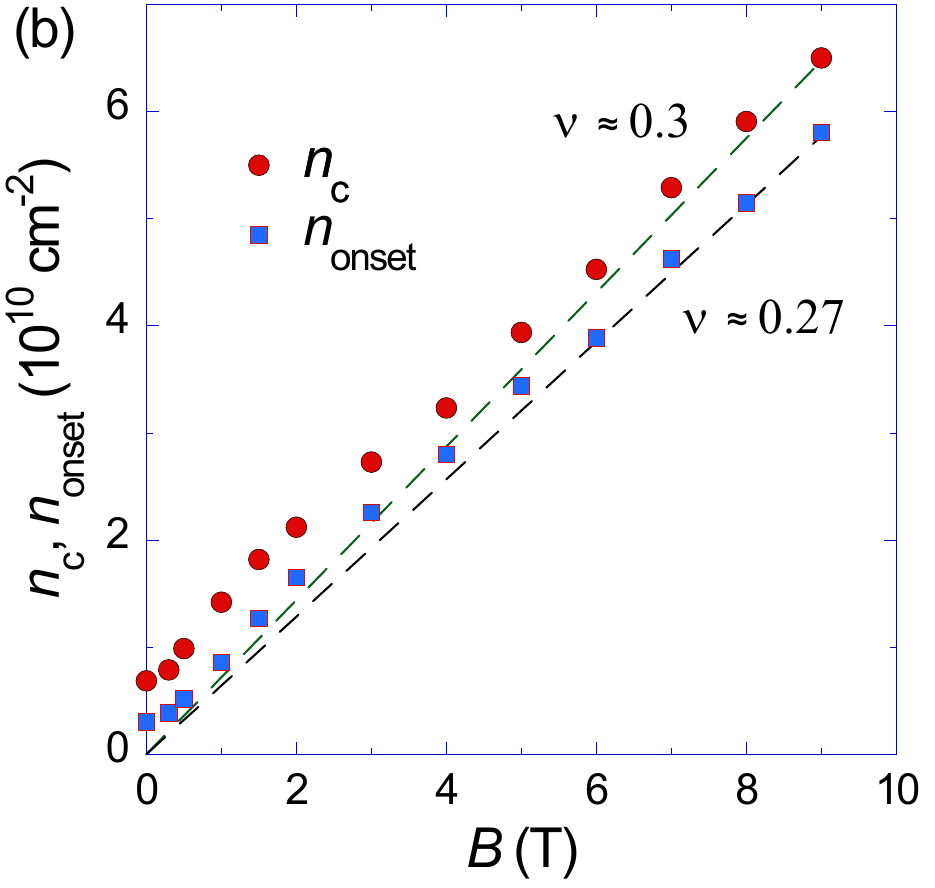}
\caption{(a) The voltage $V_{\text{onset}}$ for the onset of the double-threshold $V$-$I$ curves at $T=30$~mK as a function of the magnetic field.  The dashed line is a guide to the eye. (b) The corresponding electron density $n_{\text{onset}}$ for the onset of the double-threshold $V$-$I$ curves at $T=30$~mK as a function of the magnetic field (squares).  Also shown is the critical density $n_{\text c}$ for the metal-insulator transition versus magnetic field (circles).  The dashed lines indicate the slopes of the dependences at high $B$. From Ref.~\cite{melnikov2025stabilization}.}\label{fig9}
\end{figure}

Voltage-current characteristics at a temperature of $T=60$~mK for $n_{\text{s}}=1.67\times 10^{10}$~cm$^{-2}$ under various magnetic fields are illustrated in Fig.~\ref{fig8}(c). As the magnetic field is increased, the double-threshold voltage-current curves emerge.

In Fig.~\ref{fig9}(a), the voltage \( V_{\text{onset}} \) is shown, which is determined by linearly extrapolating the \( V \)-\( I \) curve at the onset of the double-threshold behavior at \( T = 30 \) mK  to zero current. This voltage is plotted as a function of the magnetic field. \( V_{\text{onset}} \) decreases by an order of magnitude as the magnetic field increases up to \( B \approx 2 \) T. Beyond this point, in higher magnetic fields, the value of \( V_{\text{onset}} \) continues to decrease with increasing \( B \), although at a much slower rate.

In Figure~\ref{fig9}(b), the electron density \( n_{\text{onset}} \) at the onset of the double-threshold voltage-current curves measured at \( T = 30 \, \text{mK} \) is plotted as a function of the magnetic field (represented by squares). Also included is the critical density \( n_{\text{c}} \) for the metal-insulator transition versus the magnetic field (represented by circles). This critical density is determined by the vanishing nonlinearity of the $V-I$ curves when extrapolated from the insulating phase, as detailed in Ref.~\cite{melnikov2019quantum}. Both \( n_{\text{onset}} \) and \( n_{\text{c}} \) increase with the magnetic field. Notably, for all magnetic fields considered, \( n_{\text{onset}} \) remains below \( n_{\text{c}} \). At high magnetic fields, both quantities tend to linear dependences that correspond to a filling factor of \( \nu = n_{\text{s}}hc/eB \approx 0.27 \) for \( n_{\text{onset}} \) and \( \nu \approx 0.3 \) for \( n_{\text{c}} \).
\begin{figure}
\centering
\includegraphics[width=8 cm]{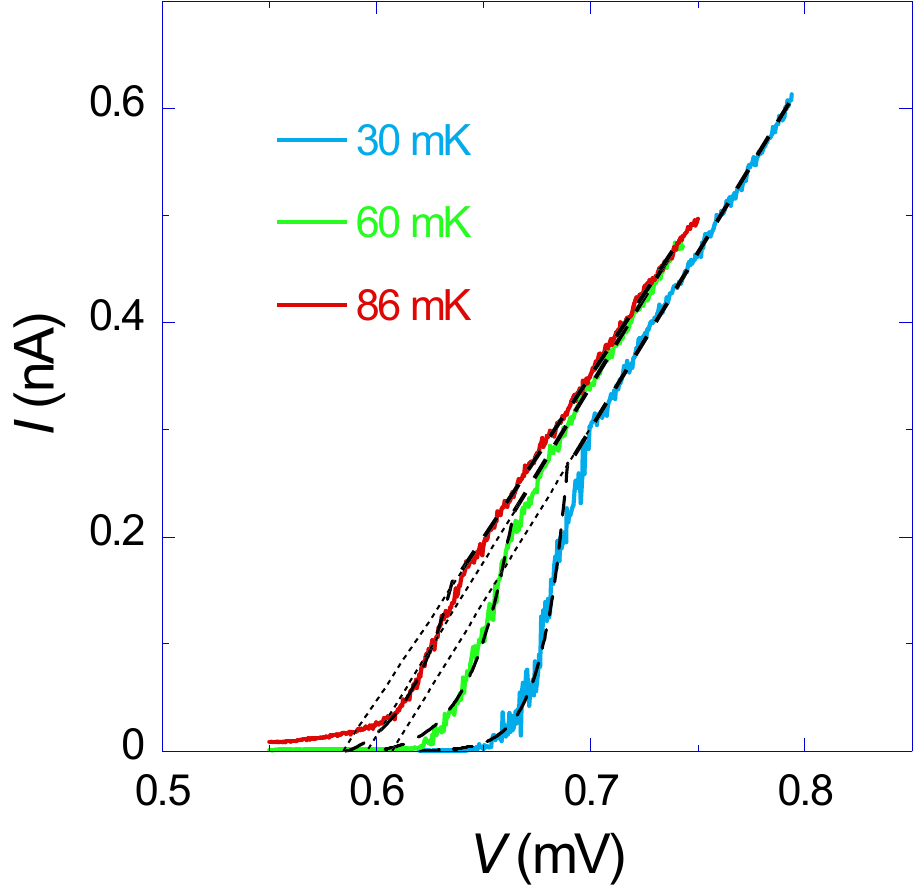}
\caption{Voltage-current characteristics at $B=4$~T and $n_{\text s}=2.72\times 10^{10}$~cm$^{-2}$ at different temperatures.  The fits corresponding to the activation energy $U_{\text c}=1.5$~K in Eq.~({\ref I}) are shown by dashed lines. From Ref.~\cite{melnikov2025stabilization}.}\label{fig10}
\end{figure}

In Fig.~\ref{fig10}, the $V$-$I$ characteristics at a magnetic field of $B=4$~T and electron density of $n_{\text s}=2.72\times 10^{10}$~cm$^{-2}$ are presented at different temperatures. The fits corresponding to the activation energy $U_{\text c}=1.5$~K accurately describe the data. Unlike the case at $B=0$, where the $V$-$I$ curves saturate below $T=60$~mK \cite{melnikov2024collective}, the $V$-$I$ curves in a strong magnetic field exhibit the expected temperature dependence down to $T=30$~mK. This indicates no overheating down to the lowest temperatures used in the experiments.

The observed increase in the onset density \( n_{\text{onset}} \) for the double-threshold \( V \)-\( I \) curves in response to increasing magnetic field
indicates that the quantum electron solid is becoming more stabilized in perpendicular magnetic fields. This finding aligns with theoretical predictions that applying a perpendicular magnetic field should encourage the formation of a Wigner solid by reducing the amplitude of zero-point vibrations of the electrons at their lattice sites \cite{lozovik1975crystallization, ulinich1978phase, fukuyama1975two, eguiluz1983two}. The corresponding filling factor \( \nu \approx 0.27 \) observed in this two-valley 2D electron system is reasonably consistent with theoretical expectations, which indicate that in a disorderless single-valley 2D electron system, the Wigner crystal should form at filling factors \( \nu \lesssim 0.15 \) \cite{lam1984liquid, levesque1984crystallization}. The intermediate insulating region between \( n_{\text{onset}}(B) \) and \( n_{\text{c}}(B) \), which precedes the formation of the electron solid, may be attributed to a fluctuation region. Additionally, the observed decrease in the onset voltage \( V_{\text{onset}} \) for the double-threshold \( V \)-\( I \) curves with increasing magnetic field also reflects the stabilizing effects of the magnetic field. Qualitatively, the behavior of \( V_{\text{onset}} \) can be explained as follows: in a zero magnetic field, the amplitude of the zero-point vibrations of the electrons at their lattice sites is relatively large, allowing for easy deformation of the electron solid near the pinning centers. Consequently, electrons should be strongly pinned by these centers, leading to a relatively high onset voltage. Conversely, in strong perpendicular magnetic fields, this relationship reverses. The increased rigidity or uniformity of the electron solid in the magnetic fields may also account for the pronounced temperature dependence of the \( V \)-\( I \) curves observed down to the lowest temperatures, which is not seen at \( B=0 \).

\subsection{Inequivalence of the low-density insulating state and quantum Hall insulating states in SiGe/Si/SiGe heterostructures}

There have been claims of the detection of the Wigner crystal at low Landau fillings in the 2D carrier system in AlGaAs/GaAs heterojunctions, based on the observation of resonance in the real part of frequency-dependent diagonal microwave conductivity that was interpreted as the pinning mode of Wigner crystal domains oscillating in the disorder potential (see, \textit{e.g.}, Refs.~\cite{williams1991conduction,engel1997microwave,ye2002correlation,sambandamurthy2006pinning,moon2014pinning,freeman2024origin}). Similar resonances were observed in AlGaAs/GaAs 2D electron systems in the quantum Hall regime near integer and fractional Landau fillings and interpreted as originating from a Wigner crystal state formed by quasi-particles with density determined by the deviation from the integer or fractional filling factor \cite{chen2003microwave,lewis2004evidence,lewis2004wigner,zhu2010observation,hatke2014microwave,moon2015microwave,kim2021the}.

\begin{figure}
\centering
\includegraphics[width=6.3 cm]{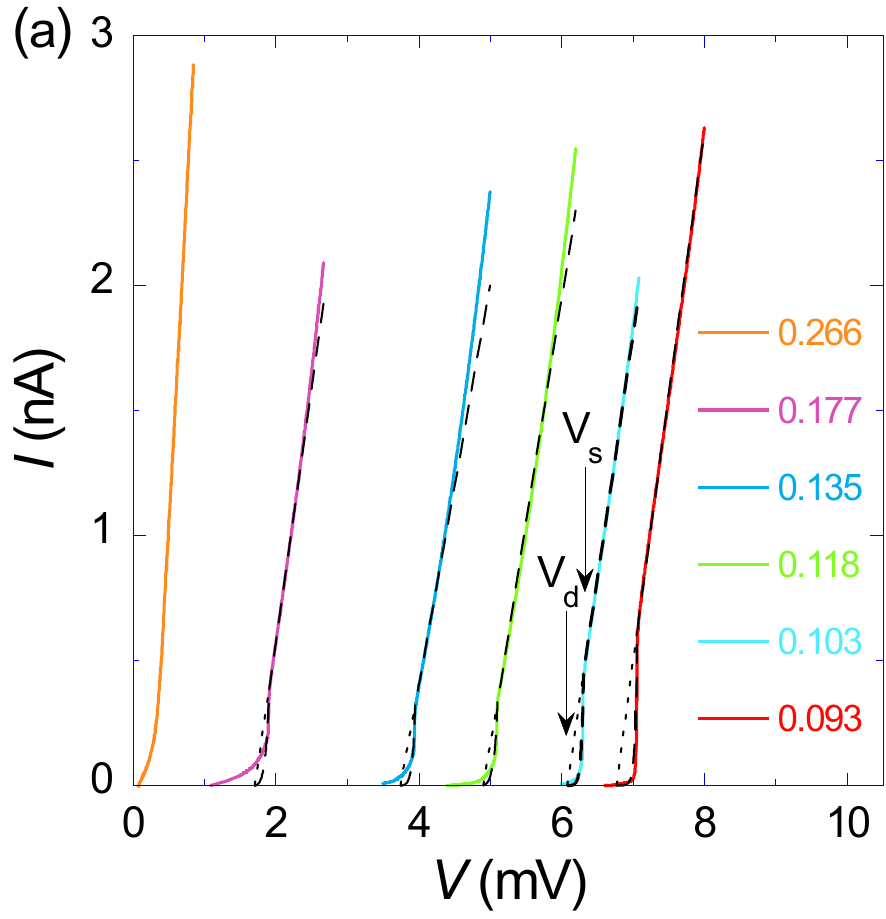}\\
\includegraphics[width=6.63 cm]{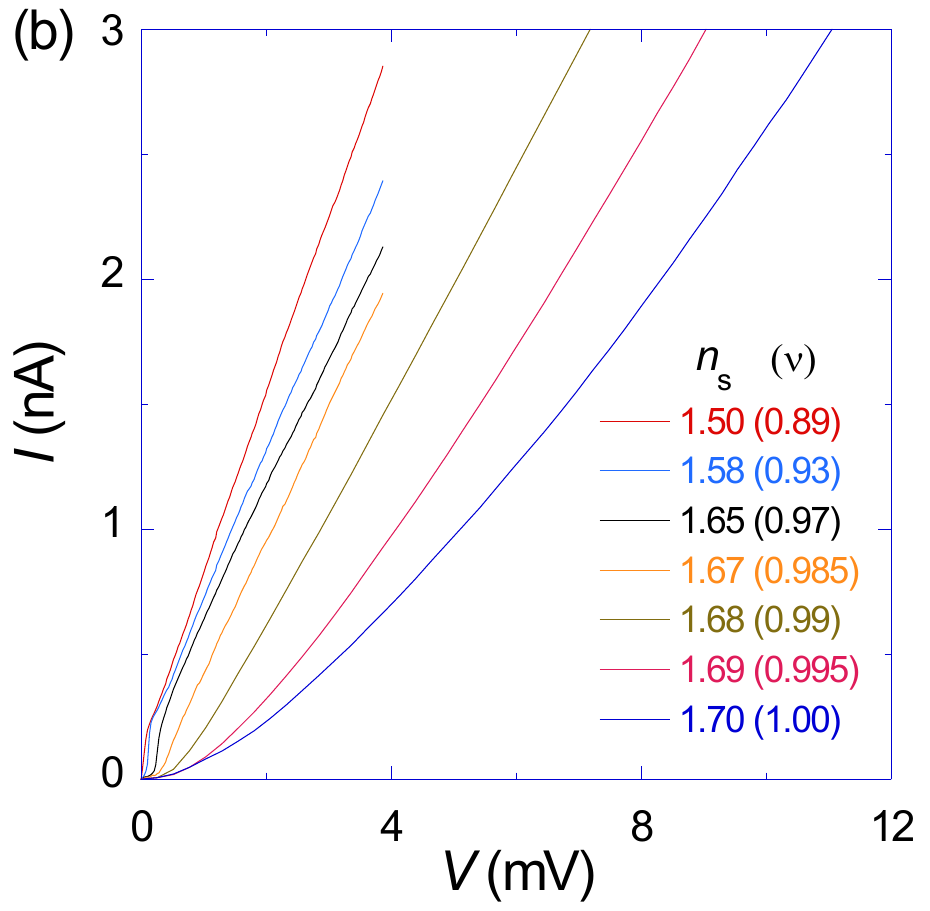}\\
\includegraphics[width=6.63 cm]{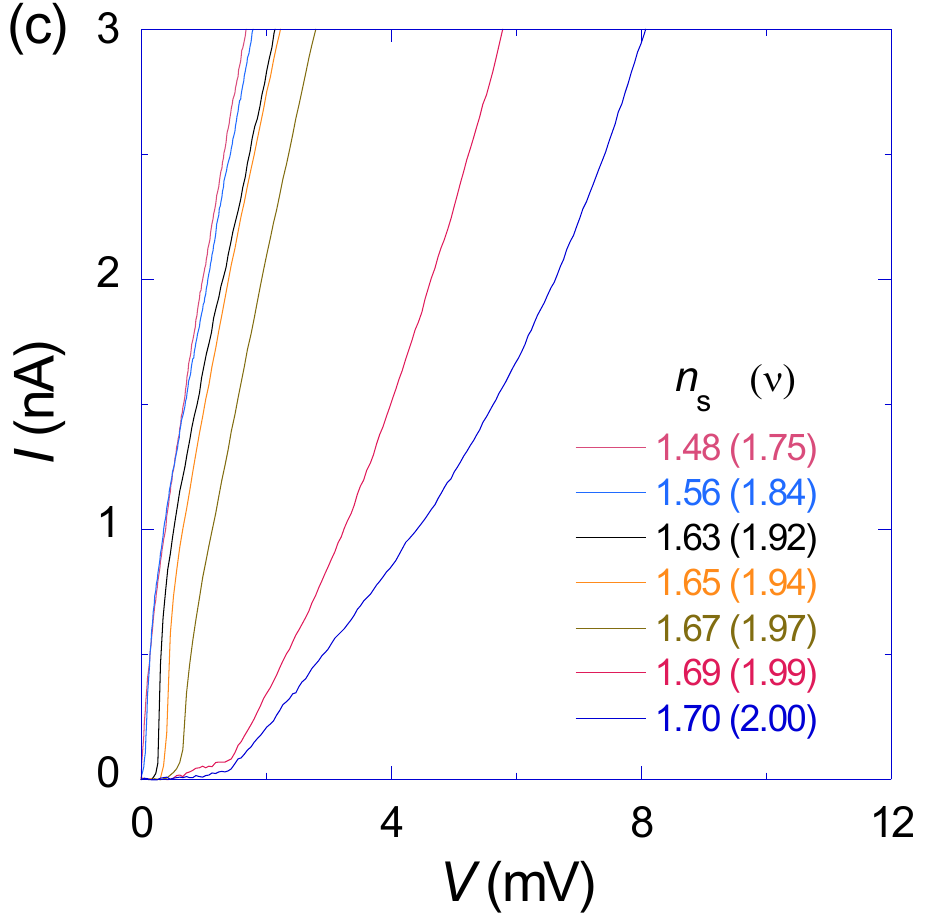}
\caption{(a) Voltage-current characteristics at different electron densities in the low-density insulating state at $B=4$~T and $T=30$~mK. The electron densities are indicated in units of $10^{11}$~cm$^{-2}$. Also shown are the dynamic threshold $V_{\text d}$ obtained by the extrapolation (dotted line) of the linear part of the $V$-$I$ curves to zero current and the static threshold $V_{\text s}$.  The dashed lines are fits to the data using Eq.~(\ref{I}). (b)~Voltage-current characteristics recalculated from the $I_{\text {sd}}$-$V_{\text {xx}}$ data at different electron densities at $T$=30 mK in $B=7$~T for $\nu\le 1$. The electron densities are indicated in units of $10^{11}$ cm$^{-2}$, along with the filling factors in brackets. (c)~Voltage-current characteristics recalculated from the $I_{\text {sd}}$-$V_{\text {xx}}$ data at different electron densities at $T$=30 mK in $B=3.5$~T for $\nu\le 2$. The electron densities are indicated in units of $10^{11}$ cm$^{-2}$, along with the filling factors in brackets. Adapted from Ref.~\cite{melnikov2025inequivalence}.}\label{fig11}
\end{figure}

The voltage-current ($V$-$I$) characteristics measured at different electron densities in the low-density insulating state at a magnetic field of $B = 4$~T and a temperature of $T = 30$~mK are shown in Fig.~\ref{fig11}(a) \cite{melnikov2025inequivalence}; here, the voltage is applied between the Hall potential probes for the configuration to be relevant for comparison with the data measured near Landau filling factors $\nu=1$ and $\nu=2$. At electron densities below $n_{\text s} \approx 0.26 \times 10^{11}$~cm$^{-2}$, two-threshold voltage-current curves are observed. As the electron density decreases and one moves further into the low-density insulating state, the double-threshold $V$-$I$ curves shift to higher voltages. This behavior is similar to what has been observed in Refs.~\cite{brussarski2018transport, melnikov2024collective, melnikov2025stabilization, shashkin2025transport} and discussed in earlier sections.

In Figs.~\ref{fig11}(b,c), the $V-I$ curves are shown, which are recalculated from breakdown dependences of the longitudinal voltage $V_{\text {xx}}$ on source-drain current $I_{\text {sd}}$ at different electron densities in a fixed magnetic field at $T=30$~mK for the quantum Hall insulating states near Landau filling factors $\nu=1$ and $\nu=2$. As long as the magnetoresistivity $\rho_{\text {xx}}$ is much smaller than the Hall resistivity $\rho_{\text {xy}}=h/\nu_0 e^2$ (here $\nu_0$ is an integer), the dissipative current $I$ between the opposite edges of the sample and the Hall voltage $V$ are related to the measured values through the quantized value of $\rho_{\text {xy}}$: $I=V_{\text {xx}}/\rho_{\text {xy}}$ and $V=I_{\text {sd}}\rho_{\text {xy}}$, where the measured value $V_{\text {xx}}$ should be normalized by the aspect ratio. The dependence $I(V)$ is a voltage-current characteristic equivalent to the Corbino geometry case \cite{shashkin1994insulating}. These characteristics look similar near $\nu=1$ and $\nu=2$. At the maximum deviations $\Delta n_{\text s}=|\nu-\nu_0|eB/hc$, the current increases sharply with increasing applied voltage, and then the slope of the $V$-$I$ curves decreases, corresponding to an approximately proportional increase of the current with voltage. As one enters the quantum Hall insulating state by reducing the deviation $\Delta n_{\text s}$, the initial steep rise of the current and the reduction in the slope of the $V$-$I$ curves disappear. The obtained $V$-$I$ curves are in contrast to the double-threshold $V$-$I$ characteristics in the low-density insulating state from Fig.~\ref{fig11}(a). It should be emphasized that the qualitative difference between the $V$-$I$ curves in the low-density insulating state and quantum Hall insulating states is observed in similar ranges of voltages, currents, and electron densities ($n_{\text s}$) or quasi-particle densities ($\Delta n_{\text s}$).

Assuming the existence of a Wigner crystal state formed by quasi-particles with density determined by the deviation from the integer filling factor, one expects similar data in the low-density insulating state and quantum Hall insulating states, according to Refs.~\cite{chen2003microwave,lewis2004evidence,lewis2004wigner,zhu2010observation,hatke2014microwave,moon2015microwave,kim2021the}. It should be emphasized that the key idea underlying this assumption is that the quasi-particles can be considered separately as an independent subsystem. However, the experimental results obtained in the 2D electron system in SiGe/Si/SiGe quantum wells do not confirm the occurrence of a quasi-particle quantum Hall Wigner solid. The low-density insulating state is characterized by the observed double-threshold voltage-current curves, which serve as a signature of the quantum Wigner solid \cite{brussarski2018transport,melnikov2024collective,melnikov2025stabilization,shashkin2025transport}. In contrast, significantly different $V$-$I$ curves are observed in the integer quantum Hall insulating states. This finding indicates that the quasi-particles near integer filling do not form an independent subsystem. A possible reason may be the mixing of Landau levels due to electron-electron interactions that strongly exceed the cyclotron energy in this 2D electron system. For comparison, the 2D electron system in AlGaAs/GaAs heterostructures, where similar data were reported in the low-density insulating state and quantum Hall insulating states, is characterized by electron-electron interactions that are comparable to the cyclotron energy due to an appreciably smaller (by a factor of about 3) effective mass, in which case the mixing of Landau levels is expected to be less relevant. This difference can be essential in interpreting the experimental results in both electron systems.

\section{Conclusions}

Two-threshold voltage-current characteristics, accompanied by a peak in broadband current noise occurring between these two threshold voltages, have been observed in the insulating regime at low electron densities in 2D electron systems within Si MOSFETs and ultra-high mobility SiGe/Si/SiGe heterostructures. These findings can be explained by a phenomenological theory of the collective depinning of elastic structures, which naturally generates a peak in broadband current noise between the dynamic and static thresholds, transitioning to the sliding of the solid over a pinning barrier above the static threshold.  The obtained results provide evidence for the formation of a quantum electron solid in these electron systems and demonstrate the generality of this effect across different classes of electron systems. Furthermore, applying a perpendicular magnetic field promotes the double-threshold behavior, allowing it to occur at voltages that are an order of magnitude lower and at significantly higher electron densities compared to the zero-field case. This indicates the stabilization of the quantum electron solid, aligning with theoretical predictions.  The double-threshold voltage-current curves, representative of electron solid formation at low densities, are not observed in the quantum Hall regime, which does not confirm the existence of a quasi-particle quantum Hall Wigner solid and indicates that quasi-particles near integer filling do not form an independent subsystem.

\section*{Disclosure statement}

The authors declare no conflict of interest.

\section*{Funding}

This investigation was supported by the RF state task.


\begin{thebibliography}{10}
\providecommand{\url}[1]{\normalfont{#1}}
\providecommand{\urlprefix}{Available from: }

\bibitem{wigner1934on}
Wigner~E. On the interaction of electrons in metals. Phys Rev. 1934;\hspace{0pt}46:1002--1011.

\bibitem{chaplik1972possible}
Chaplik~AV. Possible crystallization of charge carriers in low-density inversion layers. Sov Phys JETP. 1972;\hspace{0pt}35:395--398.

\bibitem{tanatar1989ground}
Tanatar~B, Ceperley~DM. Ground state of the two-dimensional electron gas. Phys Rev B. 1989;\hspace{0pt}39:5005--5016.

\bibitem{shashkin2001indication}
Shashkin~AA, Kravchenko~SV, Dolgopolov~VT, et~al. Indication of the ferromagnetic instability in a dilute two-dimensional electron system. Phys Rev Lett. 2001;\hspace{0pt}87:086801.

\bibitem{shashkin2002sharp}
Shashkin~AA, Kravchenko~SV, Dolgopolov~VT, et~al. Sharp increase of the effective mass near the critical density in a metallic two-dimensional electron system. Phys Rev B. 2002;\hspace{0pt}66:073303.

\bibitem{attaccalite2002correlation}
Attaccalite~C, Moroni~S, Gori-Giorgi~P, et~al. Correlation energy and spin polarization in the 2{D} electron gas. Phys Rev Lett. 2002;\hspace{0pt}88:256601.

\bibitem{spivak2004phases}
Spivak~B, Kivelson~SA. Phases intermediate between a two-dimensional electron liquid and {Wigner} crystal. Phys Rev B. 2004;\hspace{0pt}70:155114.

\bibitem{shashkin2006pauli}
Shashkin~AA, Anissimova~S, Sakr~MR, et~al. Pauli spin susceptibility of a strongly correlated two-dimensional electron liquid. Phys Rev Lett. 2006;\hspace{0pt}96:036403.

\bibitem{mokashi2012critical}
Mokashi~A, Li~S, Wen~B, et~al. Critical behavior of a strongly interacting {2D} electron system. Phys Rev Lett. 2012;\hspace{0pt}109:096405.

\bibitem{melnikov2017indication}
Melnikov~MY, Shashkin~AA, Dolgopolov~VT, et~al. Indication of band flattening at the {F}ermi level in a strongly correlated electron system. Sci Rep. 2017;\hspace{0pt}7:14539.

\bibitem{kagalovsky2020hartree}
Kagalovsky~V, Kravchenko~SV, Nemirovsky~D. Hartree-{F}ock description of a {W}igner crystal in two dimensions. Physica E. 2020;\hspace{0pt}119:114016.

\bibitem{grimes1979evidence}
Grimes~CC, Adams~G. Evidence for a liquid-to-crystal phase transition in a classical, two-dimensional sheet of electrons. Phys Rev Lett. 1979 Mar;\hspace{0pt}42:795--798.

\bibitem{andrei1988observation}
Andrei~EY, Deville~G, Glattli~DC, et~al. Observation of a magnetically induced {W}igner solid. Phys Rev Lett. 1988;\hspace{0pt}60:2765--2768.

\bibitem{williams1991conduction}
Williams~FIB, Wright~PA, Clark~RG, et~al. Conduction threshold and pinning frequency of magnetically induced {W}igner solid. Phys Rev Lett. 1991;\hspace{0pt}66:3285--3288.

\bibitem{goldman1990evidence}
Goldman~VJ, Santos~M, Shayegan~M, et~al. Evidence for two-dimensional quantum {W}igner crystal. Phys Rev Lett. 1990;\hspace{0pt}65:2189--2192.

\bibitem{jiang1990quantum}
Jiang~HW, Willett~RL, Stormer~HL, et~al. Quantum liquid versus electron solid around \ensuremath{\nu}=1/5 {L}andau-level filling. Phys Rev Lett. 1990 Jul;\hspace{0pt}65:633--636.

\bibitem{jiang1991magnetotransport}
Jiang~HW, Stormer~HL, Tsui~DC, et~al. Magnetotransport studies of the insulating phase around $\nu = 1/5$ {L}andau-level filling. Phys Rev B. 1991;\hspace{0pt}44:8107--8114.

\bibitem{diorio1992reentrant}
D'Iorio~M, Pudalov~VM, Semenchinsky~SG. Reentrant insulating phase in {S}i inversion layers in low magnetic fields. Phys Rev B. 1992;\hspace{0pt}46:15992--16004.

\bibitem{pudalov1993zero}
Pudalov~VM, D'Iorio~M, Kravchenko~SV, et~al. Zero-magnetic-field collective insulator phase in a dilute 2{D} electron system. Phys Rev Lett. 1993;\hspace{0pt}70:1866--1869.

\bibitem{giamarchi2003electronic}
Giamarchi~T. Electronic glasses. In: Quantum phenomena in mesoscopic systems. IOS Press; 2003. p. 303--339.

\bibitem{qiu2012connecting}
Qiu~RLJ, Gao~XPA, Pfeiffer~LN, et~al. Connecting the reentrant insulating phase and the zero-field metal-insulator transition in a 2{D} hole system. Phys Rev Lett. 2012;\hspace{0pt}108:106404.

\bibitem{knighton2014reentrant}
Knighton~T, Wu~Z, Tarquini~V, et~al. Reentrant insulating phases in the integer quantum {H}all regime. Phys Rev B. 2014 Oct;\hspace{0pt}90:165117.

\bibitem{qiu2018new}
Qiu~R, Liu~CW, Liu~S, et~al. New reentrant insulating phases in strongly interacting {2D} systems with low disorder. Appl Sci. 2018;\hspace{0pt}8:1909.

\bibitem{knighton2018evidence}
Knighton~T, Wu~Z, Huang~J, et~al. Evidence of two-stage melting of {W}igner solids. Phys Rev B. 2018;\hspace{0pt}97:085135.

\bibitem{falson2022competing}
Falson~J, Sodemann~I, Skinner~B, et~al. Competing correlated states around the zero-field {W}igner crystallization transition of electrons in two dimensions. Nat Mater. 2022 Mar;\hspace{0pt}21(3):311--316.

\bibitem{hossain2022anisotropic}
Hossain~MS, Ma~MK, Villegas-Rosales~KA, et~al. Anisotropic two-dimensional disordered {W}igner solid. Phys Rev Lett. 2022 Jul;\hspace{0pt}129:036601.

\bibitem{madathil2023moving}
Madathil~PT, Rosales~KAV, Chung~YJ, et~al. Moving crystal phases of a quantum {W}igner solid in an ultra-high-quality {2D} electron system. Phys Rev Lett. 2023 Dec;\hspace{0pt}131:236501.

\bibitem{marianer1992effective}
Marianer~S, Shklovskii~BI. Effective temperature of hopping electrons in a strong electric field. Phys Rev B. 1992;\hspace{0pt}46:13100--13103.

\bibitem{dolgopolov1992metal}
Dolgopolov~VT, Kravchenko~GV, Shashkin~AA, et~al. Metal-insulator transition in {Si} inversion layers in the extreme quantum limit. Phys Rev B. 1992;\hspace{0pt}46:13303--13308.

\bibitem{shashkin1994insulating}
Shashkin~AA, Dolgopolov~VT, Kravchenko~GV. Insulating phases in a two-dimensional electron system of high-mobility {Si MOSFET's}. Phys Rev B. 1994 May;\hspace{0pt}49:14486--14495.

\bibitem{shashkin2005metal}
Shashkin~AA. Metal-insulator transitions and the effects of electron-electron interactions in two-dimensional electron systems. Phys Usp. 2005 feb;\hspace{0pt}48:129--149.

\bibitem{brussarski2018transport}
Brussarski~P, Li~S, Kravchenko~SV, et~al. Transport evidence for a sliding two-dimensional quantum electron solid. Nat Commun. 2018;\hspace{0pt}9:3803.

\bibitem{heemskerk1998nonlinear}
Heemskerk~R, Klapwijk~TM. Nonlinear resistivity at the metal-insulator transition in a two-dimensional electron gas. Phys Rev B. 1998 Jul;\hspace{0pt}58:R1754--R1757.

\bibitem{lu2009observation}
Lu~TM, Tsui~DC, Lee~CH, et~al. Observation of two-dimensional electron gas in a {Si} quantum well with mobility of $1.6\times10^6$~cm$^2$/{V}s. Appl Phys Lett. 2009;\hspace{0pt}94:182102.

\bibitem{lu2010erratum}
Lu~TM, Tsui~DC, Lee~CH, et~al. Erratum: ``{O}bservation of two-dimensional electron gas in a {Si} quantum well with mobility of $1.6\times10^6$~cm$^2$/{V}s''. Appl Phys Lett. 2010;\hspace{0pt}97:059904.

\bibitem{huang2012mobility}
Huang~SH, Lu~TM, Lu~SC, et~al. Mobility enhancement of strained {Si} by optimized {SiGe/Si/SiGe} structures. Appl Phys Lett. 2012;\hspace{0pt}101(4):42111.

\bibitem{schaffler1997high}
Sch\"affler~F. High-mobility {S}i and {G}e structures. Semicond Sci Technol. 1997 dec;\hspace{0pt}12(12):1515.

\bibitem{melnikov2024triple}
Melnikov~MY, Shashkin~AA, Huang~SH, et~al. {Triple-top-gate technique for studying the strongly interacting {2D} electron systems in heterostructures}. Appl Phys Lett. 2024 10;\hspace{0pt}125(15):153102.

\bibitem{melnikov2015ultra}
Melnikov~MY, Shashkin~AA, Dolgopolov~VT, et~al. Ultra-high mobility two-dimensional electron gas in a {SiGe/Si/SiGe} quantum well. Appl Phys Lett. 2015;\hspace{0pt}106:092102.

\bibitem{melnikov2017unusual}
Melnikov~MY, Dolgopolov~VT, Shashkin~AA, et~al. Unusual anisotropy of inplane field magnetoresistance in ultra-high mobility {SiGe/Si/SiGe} quantum wells. J Appl Phys. 2017;\hspace{0pt}122:224301.

\bibitem{jaroszynski2002universal}
Jaroszy\ifmmode~\acute{n}\else \'{n}\fi{}ski~J, Popovi\ifmmode~\acute{c}\else \'{c}\fi{}~D, Klapwijk~TM. Universal behavior of the resistance noise across the metal-insulator transition in silicon inversion layers. Phys Rev Lett. 2002 Dec;\hspace{0pt}89:276401.

\bibitem{jaroszynski2004magnetic}
Jaroszy\ifmmode~\acute{n}\else \'{n}\fi{}ski~J, Popovi\ifmmode~\acute{c}\else \'{c}\fi{}~D, Klapwijk~TM. Magnetic-field dependence of the anomalous noise behavior in a two-dimensional electron system in silicon. Phys Rev Lett. 2004 Jun;\hspace{0pt}92:226403.

\bibitem{yeh1991flux}
Yeh~WJ, Kao~YH. Flux-flow noise in type-{II} superconductors. Phys Rev B. 1991 Jul;\hspace{0pt}44:360--373.

\bibitem{blatter1994vortices}
Blatter~G, Feigel'man~MY, Geshkenbein~YB, et~al. Vortices in high-temperature superconductors. Rev Mod Phys. 1994;\hspace{0pt}66:1125--1388.

\bibitem{bullard2008vortex}
Bullard~TJ, Das~J, Daquila~GL, et~al. Vortex washboard voltage noise in type-{II} superconductors. Eur Phys J B. 2008 Oct;\hspace{0pt}65(4):469.

\bibitem{shashkin2001metal}
Shashkin~AA, Kravchenko~SV, Klapwijk~TM. Metal-insulator transition in a {2D} electron gas: Equivalence of two approaches for determining the critical point. Phys Rev Lett. 2001;\hspace{0pt}87:266402.

\bibitem{melnikov2024collective}
Melnikov~MY, Shashkin~AA, Huang~SH, et~al. Collective depinning and sliding of a quantum {W}igner solid in a two-dimensional electron system. Phys Rev B. 2024 Jan;\hspace{0pt}109:L041114.

\bibitem{melnikov2025stabilization}
Melnikov~MY, Smirnov~DG, Shashkin~AA, et~al. Stabilization of a two-dimensional quantum electron solid in perpendicular magnetic fields. Phys Rev B. 2025 Jan;\hspace{0pt}111:L041301.

\bibitem{melnikov2019quantum}
Melnikov~MY, Shashkin~AA, Dolgopolov~VT, et~al. Quantum phase transition in ultrahigh mobility {SiGe/Si/SiGe} two-dimensional electron system. Phys Rev B. 2019;\hspace{0pt}99:081106(R).

\bibitem{lozovik1975crystallization}
Lozovik~YE, Yudson~VI. Crystallization of a two-dimensional electron gas in a magnetic field. JETP Lett. 1975;\hspace{0pt}22:11--12.

\bibitem{ulinich1978phase}
Ulinich~FP, Usov~NA. Phase diagram of a two-dimensional {W}igner crystal in a magnetic field. Sov Phys JETP. 1979;\hspace{0pt}49:147--150.

\bibitem{fukuyama1975two}
Fukuyama~H. Two-dimensional {W}igner crystal under magnetic field. Solid State Commun. 1975;\hspace{0pt}17(10):1323--1326.

\bibitem{eguiluz1983two}
Eguiluz~AG, Maradudin~AA, Elliott~RJ. Two-dimensional {W}igner lattice in a magnetic field and in the presence of a random array of pinning centers. Phys Rev B. 1983 Apr;\hspace{0pt}27:4933--4945.

\bibitem{lam1984liquid}
Lam~PK, Girvin~SM. Liquid-solid transition and the fractional quantum-{H}all effect. Phys Rev B. 1984 Jul;\hspace{0pt}30:473--475.

\bibitem{levesque1984crystallization}
Levesque~D, Weis~JJ, MacDonald~AH. Crystallization of the incompressible quantum-fluid state of a two-dimensional electron gas in a strong magnetic field. Phys Rev B. 1984 Jul;\hspace{0pt}30:1056--1058.

\bibitem{engel1997microwave}
Engel~LW, Li~CC, Shahar~D, et~al. Microwave resonances in low-filling insulating phase of two-dimensional electron system. Solid State Commun. 1997;\hspace{0pt}104(3):167--171.

\bibitem{ye2002correlation}
Ye~PD, Engel~LW, Tsui~DC, et~al. Correlation lengths of the {W}igner-crystal order in a two-dimensional electron system at high magnetic fields. Phys Rev Lett. 2002 Oct;\hspace{0pt}89:176802.

\bibitem{sambandamurthy2006pinning}
Sambandamurthy~G, Wang~Z, Lewis~RM, et~al. Pinning mode resonances of new phases of 2{D} electron systems in high magnetic fields. Solid State Commun. 2006;\hspace{0pt}140(2):100--106. Emergent phenomena in quantum Hall systems.

\bibitem{moon2014pinning}
Moon~BH, Engel~LW, Tsui~DC, et~al. Pinning modes of high-magnetic-field {W}igner solids with controlled alloy disorder. Phys Rev B. 2014 Feb;\hspace{0pt}89:075310.

\bibitem{freeman2024origin}
Freeman~ML, Madathil~PT, Pfeiffer~LN, et~al. Origin of pinning disorder in magnetic-field-induced {W}igner solids. Phys Rev Lett. 2024 Apr;\hspace{0pt}132:176301.

\bibitem{chen2003microwave}
Chen~Y, Lewis~RM, Engel~LW, et~al. Microwave resonance of the {2D} {W}igner crystal around integer {L}andau fillings. Phys Rev Lett. 2003 Jul;\hspace{0pt}91:016801.

\bibitem{lewis2004evidence}
Lewis~RM, Chen~Y, Engel~LW, et~al. Evidence of a first-order phase transition between {W}igner-crystal and bubble phases of {2D} electrons in higher {L}andau levels. Phys Rev Lett. 2004 Oct;\hspace{0pt}93:176808.

\bibitem{lewis2004wigner}
Lewis~RM, Chen~Y, Engel~LW, et~al. {W}igner crystallization about $\nu$=3. Physica E: Low-dimensional Systems and Nanostructures. 2004;\hspace{0pt}22(1):104--107. 15th International Conference on Electronic Properties of Two-Dimensional Systems (EP2DS-15).

\bibitem{zhu2010observation}
Zhu~H, Chen~YP, Jiang~P, et~al. Observation of a pinning mode in a {W}igner solid with $\ensuremath{\nu}=1/3$ fractional quantum {H}all excitations. Phys Rev Lett. 2010 Sep;\hspace{0pt}105:126803.

\bibitem{hatke2014microwave}
Hatke~AT, Liu~Y, Magill~BA, et~al. Microwave spectroscopic observation of distinct electron solid phases in wide quantum wells. Nat Commun. 2014 Jun;\hspace{0pt}5(1):4154.

\bibitem{moon2015microwave}
Moon~BH, Engel~LW, Tsui~DC, et~al. Microwave pinning modes near {L}andau filling $\ensuremath{\nu}=1$ in two-dimensional electron systems with alloy disorder. Phys Rev B. 2015 Jul;\hspace{0pt}92:035121.

\bibitem{kim2021the}
Kim~KS, Kivelson~SA. The quantum {Hall} effect in the absence of disorder. npj Quantum Materials. 2021 Mar;\hspace{0pt}6(1):22.

\bibitem{melnikov2025inequivalence}
Melnikov~MY, Smirnov~DG, Shashkin~AA, et~al. Inequivalence of the low-density insulating state and quantum {Hall} insulating states in a strongly correlated two-dimensional electron system. Phys Rev B. 2025 Oct;\hspace{0pt}112:165309.

\bibitem{shashkin2025transport}
Shashkin~AA, Melnikov~MY, Kravchenko~SV. Transport evidence for the quantum {W}igner solid formation in two-dimensional electron systems. Physica E: Low-dimensional Systems and Nanostructures. 2025;\hspace{0pt}168:116192.

\end{thebibliography}


\end{document}